\documentclass[11pt]{iopart}

\usepackage{graphicx}
\usepackage{subfigure}
\usepackage{float}
\usepackage{hyperref}
\usepackage[all]{hypcap}

\usepackage{placeins}

\begin{document}
\title[Lagrangian particle model for pellets and SPI fragments in tokamaks]{Lagrangian particle model for 3D simulation of pellets and SPI fragments in tokamaks}
\author{ R Samulyak$^{1,2}$, S Yuan$^1$, N Naitlho$^1$, P B Parks$^3$
\address{$^1$ Department of Applied Mathematics and Statistics, Stony Brook University, Stony Brook, NY, USA}
\address{$^2$ Computational Science Initiative, Brookhaven National Laboratory, Uptown, NY, USA}
\address{$^3$ General Atomics, San Diego, CA, USA}
\ead{roman.samulyak@stonybrook.edu}
}

\begin{abstract}
A 3D numerical model for the ablation of pellets and shattered pellet injection (SPI) fragments in tokamaks in the plasma disruption mitigation and fueling parameter space has been developed based on the Lagrangian particle code [R. Samulyak, X. Wang, H.-S. Chen, Lagrangian Particle Method for Compressible Fluid Dynamics, J. Comput. Phys., 362 (2018), 1-19]. The pellet code implements the low magnetic Reynolds number MHD equations, kinetic models for the electronic heating, a pellet surface ablation model, an equation of state that supports multiple ionization states, radiation, and a model for grad-B drift of the ablated material across the magnetic field. The Lagrangian particle algorithm is highly adaptive, capable of simulating a large number of fragments in 3D while eliminating numerical difficulties of dealing with the tokamak background plasma. The code has achieved good agreement with theory for spherically symmetric ablation flows. Axisymmetric simulations of neon and deuterium pellets in magnetic fields ranging from 1 to 6 Tesla have been compared with previous simulations using the FronTier code, and very good agreement has also been obtained. The main physics contribution of the paper is a detailed study of the influence of 3D effects, in particular grad-B drift, on pellet ablation rates and properties of ablation clouds. Smaller reductions of ablation rates in magnetic fields compared to axially symmetric simulations have been demonstrated because the ablated material is not confined to narrowing channels in the presence of grad-B drift. Contribution of various factors in the grad-B drift model has also been quantified. 
\end{abstract}

\noindent{\it Keywords\/}: pellet ablation, plasma disruption mitigation, pellet fueling, shattered pellet injection

\submitto{\NF}

\maketitle

\section{Introduction}\label{intro}

The ablation of small cryogenic pellets in thermonuclear plasma is central to the problem of fueling and plasma disruption mitigation in nuclear fusion devices of tokamak configuration. The most likely candidate for the  International Thermonuclear Experimental Reactor (ITER) plasma disruption mitigation system is the Shatterd Pellet Injection (SPI). In SPI, a large pellet, composed of a frozen mixture of neon and deuterium, is injected inside a breaker tube causing fragmentation. The plume of pellet fragment ablates in the plasma, radiates, and  induces a thermal quench. A prototype of the SPI-based disruption mitigation system has been successfully  tested on DIII-D \cite{Commaux16, Shiraki16, Shiraki_Commaux16}. The development of a successful SPI system for the much higher temperatures and magnetic fields of ITER is a challenging task that requires input from numerical simulations.

Theoretical and numerical simulation studies aimed to compute ablation rates of cryogenic pellets, needed for estimations of pellet penetration depths and deposition of ablated material in tokamaks, have been under development for decades.  The Neutral Gas Shielding model \cite{Parks_1978} and other theoretical works \cite{Felber_1979,Parks_distortion,Kuteev_1995} developed a robust theory for a single pellet that explained important experimental features. A 1D numerical model and a code HPI2, developed in \cite{Pegourie_HPI2}, was used to study the deposition of ablated material in plasma.
2D axisymmetric numerical simulations \cite{Ishizaki04} computed pellet ablation rates and described properties of ablation clouds in the hydrodynamic
approximation that ignored the magnetic field effect. Axisymmetric MHD simulations of deuterium fueling pellets \cite{Samulyak_2007,LuParksSam09}  and neon pellets in the plasma disruption mitigation parameter space \cite{Bosviel2020} computed the reduction of pellet ablation rates in magnetic fields.

The inherent  limitation of the previous simulation efforts was the use of 2D axially-symmetric or even 1D approximations.
In the presence of MHD forces, the ionization of the pellet ablation cloud by the heat flux of hot plasma electrons leads to channeling of the ablation flow along the magnetic field lines.  In 3D geometry, the curvature and grad-B drift polarization induced inside of the ionized ablated material by the $1/R$ toroidal field variation moves the ablated material across magnetic field lines and establishes a finite shielding length of the ablation cloud \cite{Rozhanskij_1994,Parks_2000,Parks_2005,Baylor_2000,Lang_1997}. To maintain the cylindrical symmetry of the ablation cloud, grad-B drift forces were ignored in  2D simulations, and an extra condition was imposed on the cloud to limit the pellet shielding length. Not only this introduced an artificial parameter into simulations but also lead to significant changes in the ablation cloud properties, as we show later, and over-estimated the influence of the magnetic field on pellet ablation rates. Needless to say, the axially symmetric approximation also prohibits direct simulation of SPI.

In this work, a fully 3D pellet / SPI fragment ablation code is developed based on the Lagrangian particle (LP) method \cite{SamWangChen2018} for hydrodynamic equations.  The Lagrangian particle method is highly optimal for the pellet ablation problem.  First, the Lagrangian particle method is highly adaptive to density changes, a critically important property for 3D simulations of the ablation of pellets and, especially, SPI fragments.  A Lagrangian treatment of the ablated material eliminates several numerical difficulties associated with the tokamak plasma background in an Eulerian hydro code. The Lagrangian particle method makes it possible to track the ablated material over very large distances. While it is not necessary for computing pellet ablation rates, this property is very useful when there is a need to compute long-scale deposition of the ablated material. Finally, the Lagrangian approach makes it much easier to extract relevant data for a multiscale coupling with tokamak-scale MHD codes, details of which will be presented in a forthcoming paper.  
  
The paper is organized as follows. In Section \ref{models}, we describe the governing hydro / MHD equations and their approximations, equation of state with multiple ionization support, kinetic electron heating, conductivity, radiation, and grad-B drift models. Section \ref{implementation} describes the numerical implementation of these 
models in the Lagrangian particle-based pellet / SPI code. Code verification tests are reported at the beginning of Section \ref{results}. We then present the  main physics  contribution of the paper: a detailed study of the influence of 3D effects, in particular grad-B drift, on pellet ablation rates and properties of ablation clouds. We complete the section on numerical results with a demonstration of code capabilities to perform resolved simulations of SPI. Finally, we conclude the paper with a summary of our results and plans for the future work.

\section{Main models and governing equations}\label{models}

\subsection{MHD in low Magnetic Reynolds number approximation}

Similar to  \cite{Samulyak_2007,Bosviel2020}, we assume that the near-field flow around the pellet is described by MHD equations in the low Magnetic Reynolds number approximation,  ${\delta B} / B \sim R_m << 1$, where $\delta B$ is the eddy current induced magnetic field. In Lagrangian coordinates, the equations are
\begin{eqnarray}\label{euler_eq}
&&\frac{d \rho}{d t}  = - \rho \nabla \mathbf{u}, \\
&&\rho \frac{d \mathbf{u}}{dt}= -\nabla P + \mathbf{J \times B}, \label{eq:momentum}\\
&&\rho \frac{de}{dt} = -P\nabla \cdot \mathbf{u} + \frac{1}{\sigma}\mathbf{J}^2 - \nabla \cdot \mathbf{q} - W_{rad},\label{eq:energy}\\
&& P = P(\rho,e), \label{eq:eos}
\end{eqnarray}
where $d/dt$ is the Lagrangian time derivative, $\mathbf{u}$, $\rho$ and $e$ are the velocity, density and specific internal energy, respectively, $P$ is the pressure, $\mathbf{B}$ is the magnetic field induction, $\mathbf{J}$ is the current density, and $\sigma$ is the fluid conductivity. The electron heat flux is represented by an external heat source $-\nabla \cdot \mathbf{q}$, the term  $W_{rad}$ describes radiation cooling, and the viscosity and heat conduction are neglected in the ablation cloud. 
In addition, the velocity field is modified by the grad-B drift model described in Section \ref{sec:gradB}.
The equation of state model (\ref{eq:eos}) that resolves atomic processes in the approximation of local thermodynamic equilibrium and the radiation cooling model are discussed in the next section.

The current density is obtained from Ohm's law
\begin{equation}
\mathbf{J}  = \sigma(-\nabla \phi + \mathbf{u \times B}),
\end{equation}
where $\phi$ is the electric potential in the cloud. For a general 3D problem, the electric potential must be found from the following Poisson equation 
\[
\nabla \sigma\nabla \phi = \nabla \cdot (\mathbf{u \times B})
\]
that follows from the charge conservation equation $\nabla\cdot  \mathbf{J}  = 0$, subject to an appropriate boundary condition.  
In this work, we assume that the ablation cloud is uniformly charged by the incoming plasma electrons, resulting in a constant value of $\phi$.

\subsection{Equation of state with atomic processes}\label{Saha1}

The degree of ionization is very low in proximity to the pellet surface and it becomes progressively higher further downstream as the temperature increases due to the energy deposition by hot plasma electrons streaming into the ablation cloud. In the portion of the pellet cloud which is sufficiently cold, dense, and collisional, we expect local thermodynamic equilibrium (LTE) conditions to prevail. As this part of the cloud plays a major role in all processes studied in this paper, we
find the distribution of ionized states by solving the following coupled system of Saha equations \cite{Zeldovich}, which depend only on local values of the temperature $T$ and mass density $\rho$:
\begin{equation}\label{Saha}
\frac{f_{m+1}f_e}{f_m} = \frac{2m_a}{\rho}\frac{u_{m+1}}{u_m}\left(\frac{2\pi m_ekT}{h^2}\right)^{\frac{3}{2}} exp \left(-\frac{I_{m+1}}{kT}\right), \quad m=1,\ldots,Z
\end{equation}
subject to the conservation conditions 
\[ 
\sum_m f_m = 1, \quad \sum_m mf_m = f_e,
\]
where $h$ is Planck's constant, $k$ is Boltzmann's constant, $m_e$ is the electron mass, $m_a$ is the atom (ion) mass, $u_m, m =1,\ldots,Z$ are known electron partition functions, $m$ is the ionization degree, $f_m$ is the corresponding fraction of $m$-times ionized ions, and $f_e$ is the electron fraction. This system together with the conservation equations of mass and charge suffice to completely determine the particle concentration fractions $f_m$ and $f_e$. Finally, we compute the thermodynamic pressure $P$ and specific internal energy $e$ as
\[
P = (1+f_e)\frac{\rho kT}{m_a}
\]
\[
e = \frac{3}{2}(1+f_e)\frac{kT}{m_a} + \frac{1}{m_a}\sum_m Q_mf_m + \frac{1}{m_a}\sum_m W_mf_m 
\]
where $Q_m = I_0 + I_1 + I_2 + \ldots + I_m$, $I_m$ are the successive ionization potentials, and $W_m$ is the electronic excitation of an m-ion.

For deuterium, the equation of state model based on the Saha equations for dissociation and ionization, developed in \cite{Samulyak_2007}, is used.

Solving the system of coupled nonlinear Saha equations at every computational node at every time step in a hydrodynamic code is prohibitively 
expensive. We created tabulated data sets of thermodynamic functions on a fine density - pressure mesh with logarithmic step along the density axis and used table look-up and cubic spline interpolation algorithms during runtime. Table look-up adds very little overhead
compared to the use of analytic formulas for the ideal gas.

\subsection{Plasma electron heat flux}

The electron heat flux model is similar to the one described in \cite{Ishizaki04,Samulyak_2007}, but it contains several improvements \cite{Parks20} mostly relevant to high-Z elements. The 3-D linearized Fokker-Planck kinetic equation is solved for the electron distribution function
$f(E,\mu,z)$ where $E$, $\mu$, $z$ are the energy, velocity space
variables and cosine pitch-angle with respect to the magnetic field,
respectively. 

The heat source $-\nabla \cdot q$ coming from the energy deposition by hot, long mean-free path electrons streaming into the
ablation cloud along the magnetic field lines is
\begin{equation}
-\nabla \cdot \mathbf{q} = \frac{q_\infty n_e(r,z)}{\tau_{eff}}[g(u_+) + g(u_-)],
\end{equation}
where
\[
q_\infty = \sqrt{\frac{2}{\pi m_e}}n_{eff}(kT_{e\infty})^\frac{3}{2},
\]
$n_{eff}$ is the effective plasma electron density due to the electrostatic shielding / albedo effect,
\[
n_{eff} = (1-0.001\:A)\:e^{-\Phi}\:n_{e_\infty}, 
\]
where $e^{-\Phi}$ is the decrease due to the electrostatic shielding and $A$ is the surface reflectivity due to collisional backscattering,
\[
A = 23.92\:\ln\left(1+0.2014(1+Z_*)\right).
\]
$g(u) = u^\frac{1}{2}K_1(u^\frac{1}{2})/4$, where $K_1$ is the standard modified Bessel function of the second kind and $T_{e\infty}$ is the temperature of the plasma electrons. The quantity $u_\pm = \frac{\tau_\pm}{\tau_{eff}}$ is a dimensionless opacity, where the respective line integrated densities of the ablation electrons (bound and free) are,
\[
\tau_+(r,x) = \int_{-\infty}^{x} n_e(r,x') dx' \quad and \quad \tau_-(r,x) = \int_{x}^{\infty} n_e(r,x') dx'
\]
For the special case of the spherically symmetric approximation,
\[
\tau = \int_{\infty}^{r} n_e(r') dr'.
\]

Hot electron energy flux is attenuated by a combination of slowing down and pitch angle scattering with an effective energy flux attenuation thickness given by
\[
\tau_{eff} = \tau_\infty \frac{1}{0.625 + 0.55\sqrt{1+Z_*} }, \quad \tau_\infty = \frac{T_{e\infty}^2}{8\pi e^4 \ln\Lambda},
\]
where $e$ is the elementary charge of electron. The Coulomb logarithm $\ln\Lambda$ here pertains to inelastic scattering of fast electrons off atomic (bound) electrons in the neutral gas target,
\[
\ln\Lambda = \ln\left(\frac{E}{I^*}\sqrt{\frac{e}{2}}\right),
\]
where $e$ is Napier's constant (not to be confused with elementary charge $e$ in the equation for $\tau_\infty$), and $I^*$  is the mean excitation energy for neutral atoms. The Coulomb logarithm is evaluated at energy $E \approx 2T_{e\infty}$ since that is the average energy per particle in the cloud from the distribution of semi-isotropic incident Maxwellian electrons. Finally, the heat deposition on the surface of the pellet is given by
\begin{equation} \label{qpm}
q_\pm = q_\infty \frac{1}{2}u_\pm K_2(u_\pm^\frac{1}{2}).
\end{equation}

\subsection{Transverse conductivity and radiation}

The conductivity model for hydrogenic species has been derived in ~\cite{Samulyak_2007} and its modification for high-Z materials was formulated in~\cite{Parks_SCIDAC}. Including the effects of electron-ion and electron-neutral collision, the conductivity transverse to the magnetic
field is
\begin{equation}\label{cond}
\sigma_\perp = \frac{9.7 \times 10^3 \thinspace T^\frac{3}{2}}{Zln\Lambda + 0.00443T^{2.245} \frac{n_0}{n_e}},
\end{equation}
where $n_0$ and $n_e$ are the particle densities of neutral and gas electrons, respectively. The fraction $\frac{n_0}{n_e}$ is obtained by solving the Saha system in Equation \ref{Saha}. $Z_{eff}$ is the following modified average charge state
\begin{equation}\label{eq:Zeff}
Z_{eff} = \frac{\sum\limits_{j=0}^{Z}Z_j^2\:n_j}{\sum\limits_{j=0}^{Z}Z_j\:n_j}.
\end{equation}
In the absence of neutrals, Equation \ref{cond} reduces to the Spitzer conductivity
\[
\sigma_\perp \rightarrow \sigma_\perp ^ S = \frac{9.7 \times 10^3 \thinspace T_{e}^\frac{3}{2}}{Zln\Lambda}
\]

The photon mean free path in the ablation channel is much longer compared to the channel diameter and length. The exception is the narrow region near the pellet surface, but the radiation coming from this region is very low. Therefore, a non-LTE radiation model in the thin optical limit is a good approximation. The radiation model implemented in the Lagrangian particle code is based on  tabulated data of the radiation power density $\mathbf{Q}_{rad}$ obtained using the CRETIN \cite{cretin} code.

\subsection{Pellet surface ablation model}

Since the dynamics of the pellet ablation is mostly defined by the processes in the ablation cloud, 
we use a simplified model for the cryogenic phase transition on the pellet surface which neglects some thermodynamic details of the phase
transition problem. 

Following \cite{Samulyak_2007}, we assume that all electron energy that reaches the pellet surface is completely used for vaporization of the pellet material.
The ablation on the pellet surface satisfies three boundary conditions. First, the heat diffusion in the solid pellet is slow
compared to the ablation process, therefore the pellet surface temperature is assumed constant.
Second, with the constant pellet density $\rho_{pel}$, the normal velocity of the ablated material at the surface is determined by the heat flux
into the pellet,  $q$, and the sublimation energy $\epsilon_s$:
\[
\frac{q}{\epsilon_s} =  \rho_{v}u,
\]
where $\rho_{v}$ is the vapor density at the pellet surface. The third condition is the characteristic hydrodynamic equation along the normal direction 
from the ablation cloud onto the pellet surface:
\[
\frac{\partial p}{\partial t}-c\frac{\partial p}{\partial n} - \rho c \left( \frac{\partial u}{\partial t} - c \frac{\partial u}{\partial n} \right) = (\gamma-1)\frac{\partial q_\pm}{\partial z},
\]
where c is the sound speed in the cloud, $\gamma$ is the ratio of specific heats, and z is the direction of the electron flux.

\subsection{Grad-B drift of ablated material}
\label{sec:gradB}

\begin{figure}[H]
\centering
\includegraphics[height=3in]{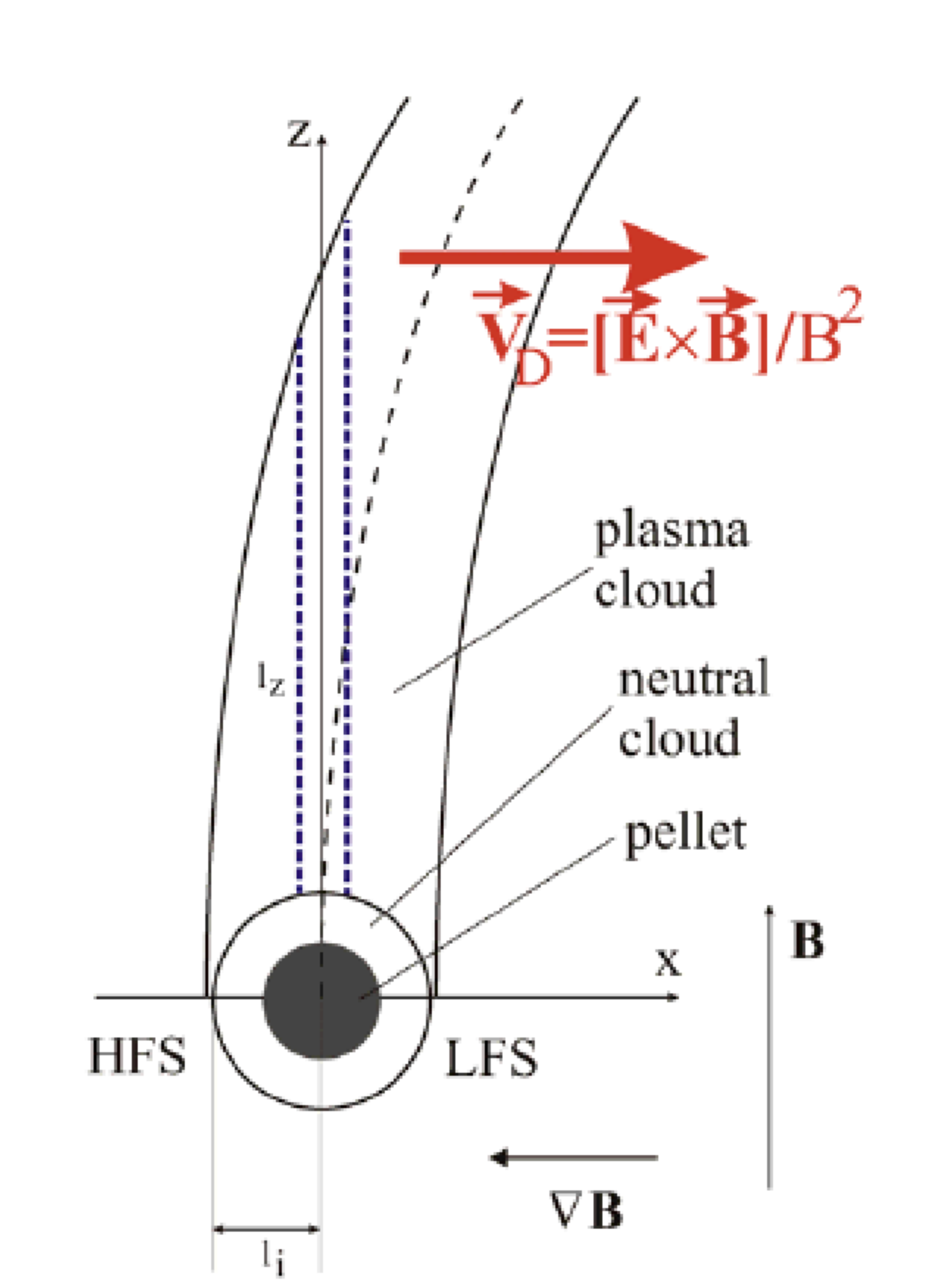}
\caption{Schematic of physics mechanisms of the grad-B induced pellet cloud drift.}
\label{drift_image}
\end{figure}

During previous pellet injection experiments, a rapid movement of the pellet ablation material towards the outward major radius $R$ direction has been observed \cite{Baylor_2000,Lang_1997}. This motion has been attributed to a vertical curvature and grad-B drift polarization induced inside the ionized ablated material by the $1/R$ toroidal field variation. The uncompensated polarization drift current inside the cloud causes charge separation at the boundary. The resulting electrostatic field induces the $\mathbf{E \times B}$ drift to the large-$R$ side of the torus as shown in Figure \ref{drift_image}.

Since the drift velocity is independent of the longitudinal coordinate (the electrostatic potential is always assumed to be uniform along the magnetic field lines), the
equation for the transverse $\mathbf{E \times B}$ drift velocity $v_D$ in the large-R direction is
governed by the following equation \cite{Parks_2005}:
\begin{equation}\label{drift_eq}
\frac{dv_D}{dt} = \frac{2 <P(1+\frac{M_\parallel^2}{2}) - P_\infty>}{R<\rho>} - u_D\frac{2B_{||}^2}{\mu_0 v_A <\rho>}.
\label{eq:drift}
\end{equation}
Here $< A >$ is the integral of quantity $A$ along a magnetic field line, $< A > \equiv \int_{-L}^{L} Adz$, where $L$ is the cloud length, $R$ denotes the tokamak major radius, $P_\infty$ is the pressure of the ambient plasma, and $<P - P_\infty>$ is called the drive integral, used in \cite{Parks_2000,Rozhanskij_1994,Pegourie_HPI2}. This drive integral was updated in  \cite{Parks_2005} to include the effect of the centrifugal force arising from the parallel flow velocity with the Mach number $M_\parallel$ in the curved toroidal magnetic field. The last terms in (\ref{drift_eq}) is the drag associated with the generation of the Alfven wave by the ablation cloud, where $v_A$ is the Alfven velocity.

\section{Numerical implementation}\label{implementation}

\subsection{Lagrangian particle method}

The pellet / SPI fragment ablation code is based on the Lagrangian particle (LP) method \cite{SamWangChen2018} for hydrodynamic equations.  The choice of a particle-based Lagrangian method is motivated by several considerations.  First, the Lagrangian particle method is highly adaptive to density changes, a critically important property for 3D simulations of the ablation of pellets and, especially, SPI fragments.  A Lagrangian treatment of the ablated material is desirable as it eliminates several numerical difficulties associated with the tokamak plasma background in an Eulerian hydro code. Finally, the Lagrangian approach makes it much easier to extract relevant data for a multiscale coupling with tokamak-scale MHD codes, details of which will be presented in a forthcoming paper.

Similar to smoothed particle hydrodynamics (SPH), the LP method represents fluid cells with Lagrangian particles and is suitable for the simulation of complex free surface / multiphase flows. The main benefits of the LP method, which is different from SPH in all other aspects, are (a) significant improvement of accuracy and mathematical rigor of the traditional SPH method, in particular the discretization of differential operators in LP is based on a polynomial fit via  weighted least squares approximation and converges  to a prescribed order, (b) robust second-order hyperbolic PDE algorithm with a choice of limiters, generalizable to higher order methods, providing accuracy and long term stability, and (c) more accurate resolution of entropy discontinuities and  states at free interfaces. Numerical stencils in the LP code include closest neighbors of every particle, properly distributed in space to minimize errors in numerical approximations of spatial derivatives. These neighborhoods are efficiently computed by constructing and searching an octree data structure containing particle information. 
The Lagrangian particle code, optimized for massively parallel supercomputers, uses p4est ("parallel forest of K-trees" ) \cite{BursteddeWilcoxGhattas11}, a parallel library that implements
a dynamic management of a collection of adaptive K-trees on distributed memory supercomputers. It has the functionality of building, refining, coarsening, 2:1 balancing, and partitioning of computational domains composed of multiple connected two-dimensional quadtrees or three-dimensional octrees, referred to as a forest of K-trees. The Lagrangian particle method is especially suitable for multiphase problems and problems with matter occupying sparse regions in space.
Some examples include high-power accelerator targets and magneto-inertial fusion applications where it offers advantages over previous simulations performed
with grid-based methods \cite{mu_target,ShihSam19}. 

The LP-based pellet / SPI fragment ablation code implements MHD equation in the low magnetic Reynolds number approximation and ablation physics models
described above. Some details of algorithmic specific to the pellet / SPI problem are presented in the next sections \ref{heatflux} and \ref{implementation_gradB}.

\subsection{Implementation of heat flux} \label{heatflux}

\begin{figure}[H]
\centering
\subfigure[]{\includegraphics[width=4in]{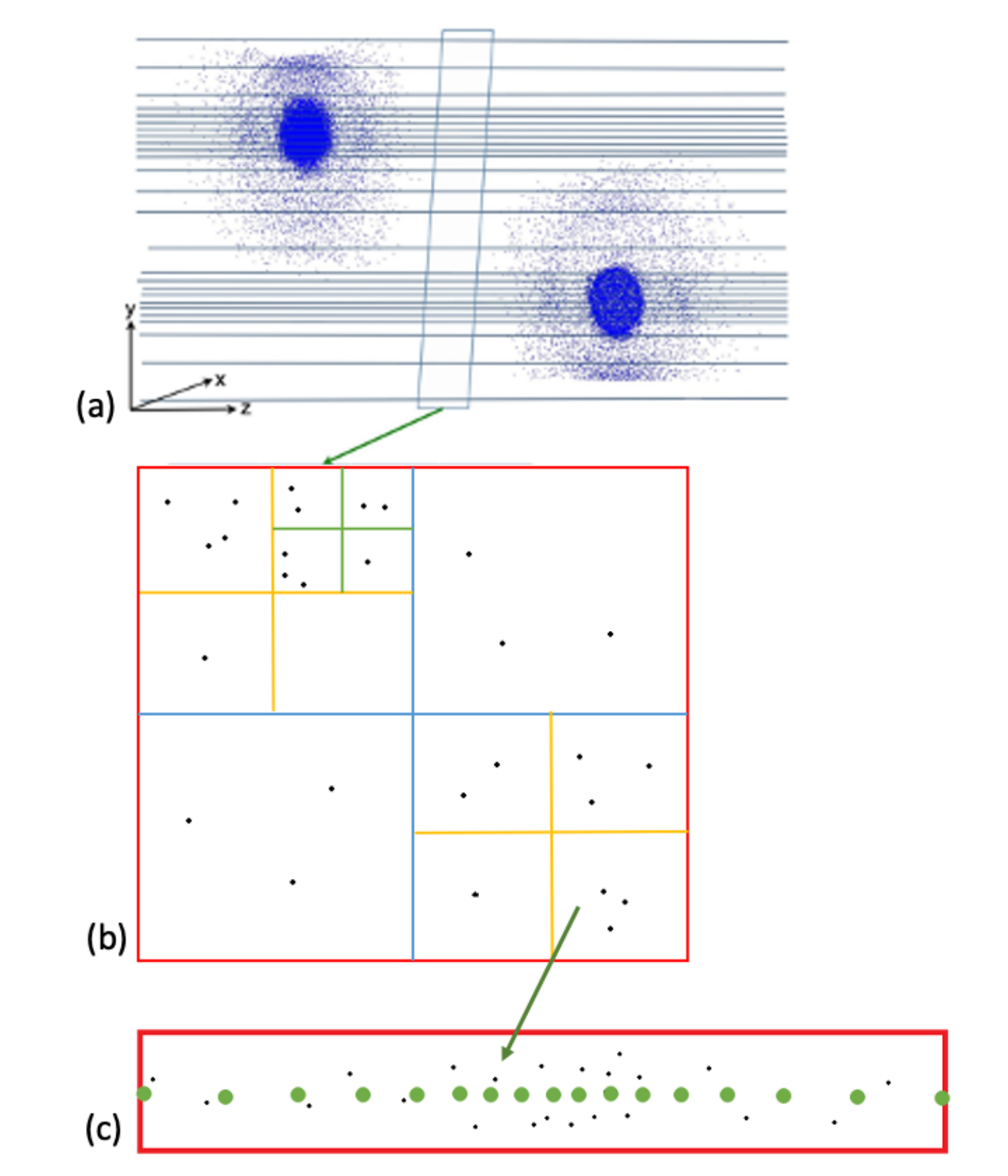}}
\caption{Schematic of line integration algorithms for kinetic models in the Lagrangian particle pellet code. (a) Integration paths are selected adaptively with their density increasing towards every SPI fragment center. (b) Quadtree is built for all particles projected to the transverse plane. (c) Particles in every quadtree cell are lifted to 3D and shown along the longitudinal direction with small dots representing
Lagrangian particles and large green dots representing computational nodes. The distance between nodes is proportional to local inter-particle distance. For clarity, very few Lagrangian particles are used in parts (b) and (c) of the schematic.}
\label{fig:int_lines}
\end{figure}

Kinetic models for the electron heat flux and models for the grad-B drift require accurate integration of ablation cloud quantities (density, pressure, etc) along magnetic fields lines. The main challenge of implementing these models is due to the particle discretization nature of the LP method and a very nonuniform distribution of particles in the ablation cloud. While most results presented in this paper are related to the ablation of single pellets, all algorithms in the LP code are designed for multiple SPI fragments. As ablation cloudlets surrounding each fragment are very dense, the selection of integration paths parallel to the magnetic field lines must also be performed adaptively, with the density of paths increasing towards every SPI fragment center (see a schematic in Figure \ref{fig:int_lines}(a)).

The integration algorithm proceeds as follows:
\begin{itemize}
\item In order to select integration paths, the entire set of particles is projected onto a plane transverse to the magnetic field, and a quadtree is constructed and refined until the number of particles in each cell does not exceed a prescribed value (Figure \ref{fig:int_lines}(b)). Each quadtree cell
specifies an integral path along the magnetic field direction and particles located in that quadtree cell are considered to be on the integral path. This algorithm ensures proper density of the integration paths for any distribution of ablating fragments. 

\item The particles in each quadtree cell are lifted back to 3D. In order to compute an integral of particle-based quantity (e.g. density for the electron heating model) along a magnetic field line passing through the center of the corresponding quadtree cell, we  sample computational nodes along each integral path (see Figure \ref{fig:int_lines}(c)). 
The purpose of computational nodes is to reduce the statistical noise in numerical integration and avoid under-resolving the density gradient. 
The distribution of nodes is such that  the distance between two adjacent nodes is proportional to the local inter-particle distance between Lagrangian particles. 

 \item For each computational node, closest neighbors among Lagrangian particles are selected using the octree data structure, and the weighted average of density is assigned to each node:
\begin{eqnarray}
    &&\rho = \sum \frac{w_{i}}{W}\rho_{i}, \nonumber \\
    &&w_{i} = \exp\left(\frac{-d^{2}}{2b^2}\right), \\
    &&W = \sum w_{i} \nonumber
\end{eqnarray}
where $\rho_{i}$ is density of the Lagrangian particle $i$, $d$ is the distance between neighbor particles and computational nodes, and $b$ is the kernel length of weight function. We select $b$ to be $\frac{1}{3}$ of the searching radius of computational nodes. 

\item Numerical integral of density along each integral path is computed using values on computational nodes. The trapezoidal method which achieves second order accuracy, consistent with the overall accuracy of the LP method, is typically used but the method easily extends to higher orders. The integration is performed and stored in both longitudinal directions from the outer domain boundaries to the present node.

\item Density integral values on every Lagrangian particle are obtained via interpolation of the corresponding quantities on computational nodes.
\end{itemize}

The algorithm has been verified using a prescribed distribution of particles and a very good agreement with the analytic solution for density integrals has been obtained.

\subsection{Grad-B Drift}\label{implementation_gradB}

The implementation of grad-B drift model  takes advantage of line integral algorithm introduced in section \ref{heatflux}. The same quadtree structure is used again for selecting integral paths for $P(1+\frac{M_\parallel^2}{2}) - P_\infty$ and $\rho$ in Equation (\ref{drift_eq}). In addition to computing the weighted average density of computational nodes, $P$ and $M$ values on nodes are assigned by the same algorithm. For the grad-B drift model, the numerical integration must be computed for computational nodes from $-\infty$ (left end) to $\infty$ (right end) along the integral paths. Assuming that computational nodes have longitudinal coordinates $x_{l}$, the values of one-sided integrals at nodes are $\int_{-\infty}^{x_{l}} Adz$. Thus, the right end node in each quadtree cell gives the correct value of $<A>$. Particles in the same quadtree cells move towards the large-R side of the torus, driven by drift acceleration from the corresponding right end nodes, independent of their longitudinal coordinates. The drift is applied only to ionized particles; neutrals do not contribute to  the integrals $<A>$ and do not experience the drift.

\section{Results}\label{results}

In this section, we present  3D Lagrangian Particle code  simulations of the ablation of neon and deuterium ($D_2$) pellets and hydrodynamic/MHD processes in the ablation cloud. The main emphasis is on the study of the ablation process in realistic tokamak magnetic fields with curvature-induced grad-B drift. But simplified approximation are also presented for the code verification purpose. In simulations presented in this paper, we keep the pellet size constant in order to compute steady-state ablation rates but algorithms for pellets decreasing in size due to the ablation process are fully implemented in the code. In the last section, we demonstrate the capability of the code to simulate the ablation of SPI fragments. As the ablation of SPI will be the subject of our forthcoming paper, the detailed analysis of SPI physics is omitted.

Most  simulations presented in this paper use the following parameters for the background plasma:  plasma electron temperature $T_{e\infty}$ = 2 keV and plasma number density $n_{e\infty} = 1 \times 10^{14} cm^{-3}$. Accounting for the effect of electrostatic shielding, we decrease the plasma number density to get $n_{eff} = 1.20491 \times 10^{13} cm^{-3}$ for Neon and $n_{eff} = 1.37944 \times 10^{13} cm^{-3}$ for $D_2$. We will call these parameters the canonical plasma parameters in the rest of the paper. Other initial simulation settings will be described as necessary.

\subsection{Code verification: hydrodynamic simulations with spherically symmetric initial conditions and sources}\label{ss}

In this section, we report results of verification simulations using spherically-symmetric initial conditions and heat sources for single neon pellets. We perform fully 3D simulations with a spherically symmetric source for the Maxwellian electron heat flux but without explicitly enforcing the symmetry of the flow field. 
The term "hydrodynamic simulations" is used to emphasize the fact that the electromagnetic terms  in the governing equations (\ref{eq:momentum} - \ref{eq:energy}) 
are ignored. Results are compared with the updated spherically symmetric transonic pellet flow model \cite{Parks_private_com} that improves approximations of the Neutral Gas Shielding (NGS) model \cite{Parks_1978}.
Verification simulations also ignore atomic processes in the ablation cloud by using the ideal gas EOS with $\gamma$ = $5/3$. We use a $40 cm \times 40 cm \times 40cm$ computational domain. Particles that leave this domain are discarded and the outflow boundary conditions are used. 

Table \ref{table_ss} summarizes results for the ablation rate of neon pellets with radii 2 mm and 5 mm, and  background plasma 
temperatures  2 keV and 5 keV, and compares it with theory. The agreement is very good, considering that simulations were performed in 3D without enforcing the 
spherical symmetry. In \cite{Bosviel2020},  we compared 1D spherically symmetric simulations performed with the FronTier code with theory in a wide range of pellet 
radii and  background plasma parameters, and the errors were typically in the range from 0.3\% to 1.1\%.  

\begin{table}[ht]
\centering
\begin{tabular}{|c|c|c|c|c|}
\hline
$r_p$, mm & $T_{\infty}$, keV & G (theory), g/s & G (LP), g/s & Error\\
\hline
2 & 2  & 64.9295  & 65.5  & +0.88\% \\
\hline
5 & 2 & 222.206 & 220 & -0.99 \% \\
\hline
2 & 5  & 297.469 & 303 & +1.87 \% \\
\hline
5 & 5 & 1019.02 & 1002 & -1.67\% \\
\hline
\end{tabular}
\caption{Comparison of ablation rates (G) in LP simulations using spherical sources with semi-analytic model.}
\label{table_ss}
\end{table}

\subsection{Code comparison: MHD simulations with axially-symmetric initial conditions}\label{ss}

In the next phase of our code verification program, we compare 3D MHD simulations of the Lagrangian particle code using axially-symmetric initial conditions and real equation of state with ionization with 2D 
cylindrically symmetric simulations obtained using the FronTier pellet code \cite{Bosviel2020,Samulyak_2007}, a grid-based code that explicitly tracks material interfaces, in particular the pellet ablation surface and the interface between the ablated material and the 
background plasma.  As in the case of spherically-symmetric simulations described above, the axial symmetry of the LP code was not strictly enforced, but the analysis of data shows that it is maintained with high accuracy. For instance, we verified that  the azimuthal component of velocity is negligibly small.
Results for both neon and deuterium fueling pellets are discussed in this section.
In the presence of MHD forces, ionization of the pellet ablation cloud by the electronic heat flux leads to the channeling of the ablation flow along the magnetic field lines. To maintain the cylindrical symmetry of the ablation cloud, grad-B drift forces (\ref{eq:drift}) are ignored, and an extra condition must be imposed on the cloud to limit the pellet shielding length. In the presence of grad-B drift, the ablated material expands along magnetic field lines and drifts across these lines
in the direction of the tokamak major radius $R$. This drift establishes a finite pellet shielding length of the ablation cloud (defined here as the extent of the ablation 
cloud from the pellet surface along a magnetic field line passing through the pellet center, as shown in Figure \ref{fig:neon_drift}(a)). Without grad-B drift, the ablated material would continue its expansion along magnetic field lines, increasing the pellet shielding, and eventually stopping the ablation process. To avoid this 
unphysical process and enable the pellet ablation to reach steady-state, we impose a finite shielding length of 16 cm,  consistent with \cite{Bosviel2020}. It is obtained by a theoretical estimate  and used for all simulations presented in this section. The finite shielding is implemented by limiting the computational domain to 32 cm in the longitudinal direction (the pellet is in the center of this domain). The Lagrangian particles that cross that boundary are discarded but only after they supply their states to numerical stencils (neighborhoods) of particles that approach the boundary within about two inter-particle distances, thus providing the outflow boundary conditions.

\begin{figure}[H]
\centering
\subfigure[Temperature (eV) distribution in 3D cloud]{\includegraphics[width=0.85\columnwidth]{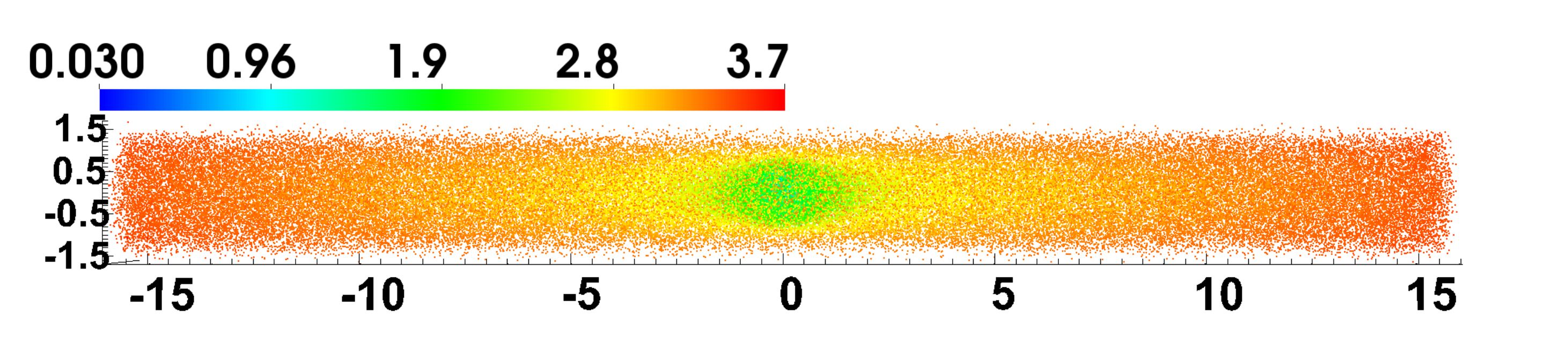}}
\subfigure[Temperature (eV) distribution on 2D slice]{\includegraphics[width=0.8\columnwidth]{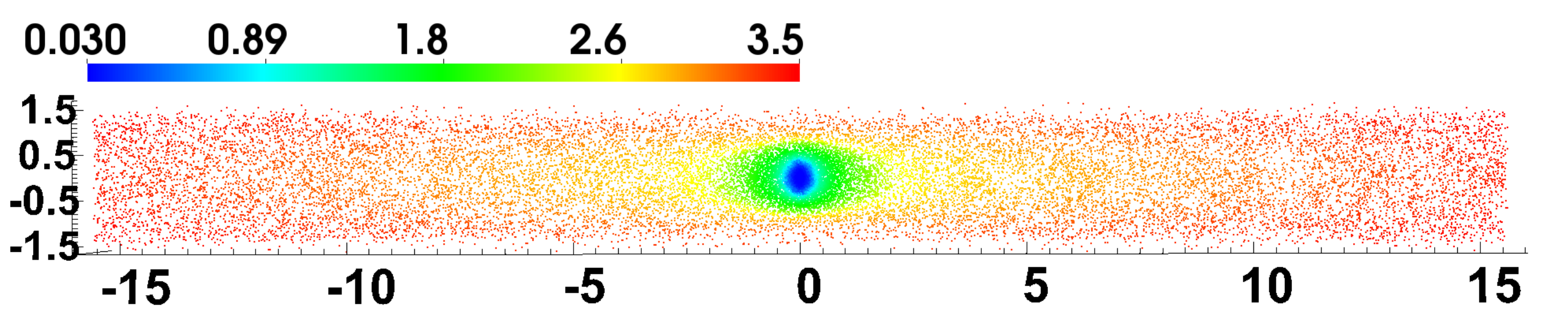}}
\subfigure[Density (g/cc)  distribution on 2D slice]{\includegraphics[width=0.8\columnwidth]{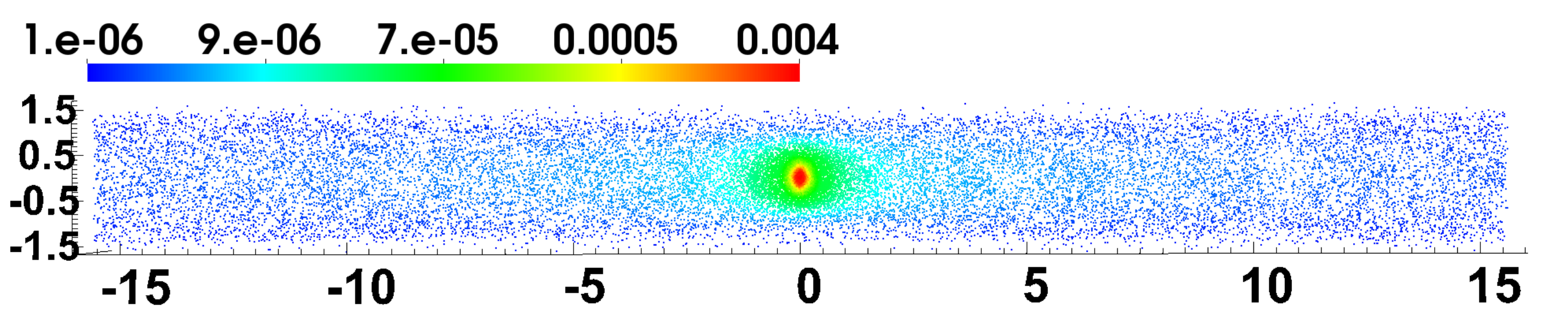}}
\subfigure[Velocity (m/s)  distribution on 2D slice]{\includegraphics[width=0.8\columnwidth]{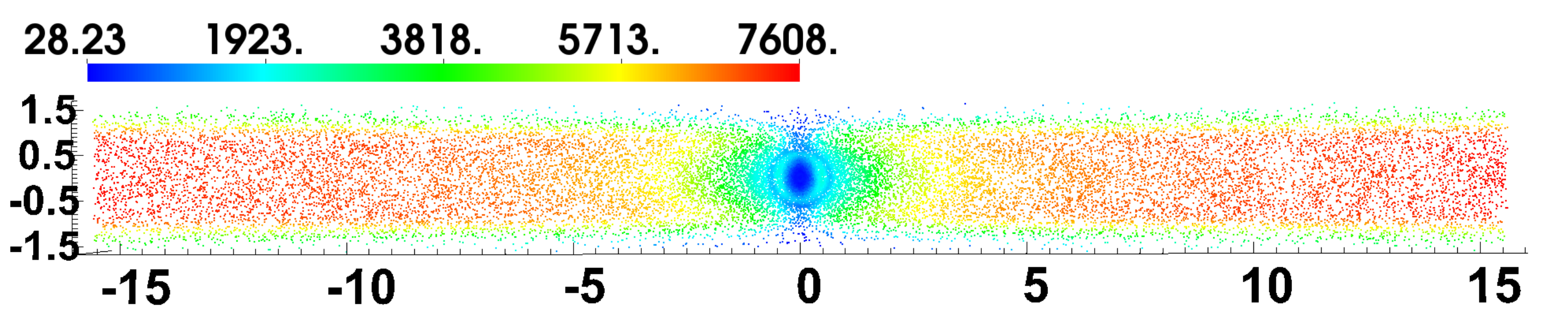}}
\subfigure[Mach number  distribution on 2D slice]{\includegraphics[width=0.8\columnwidth]{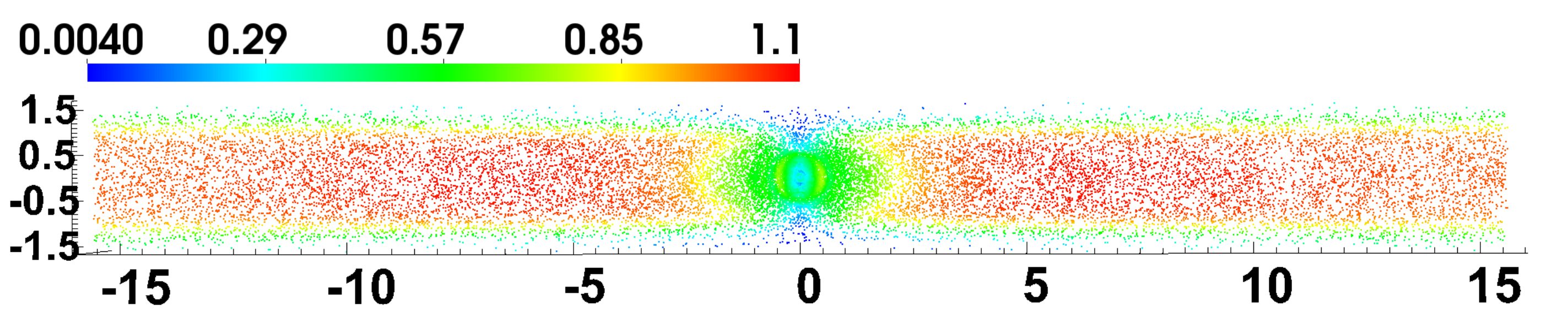}}
\caption{Distribution of ablation cloud temperature (a,b), density (c), velocity (d), and Mach number (e) at steady state  for neon pellet in B = 2 T field using fixed shielding length. 
Horizontal axes (cm) are along the magnetic field, and vertical axes (cm) are transverse to the magnetic field.}
\label{lp_fixed}
\end{figure}

We present first the overall pellet ablation process as obtained by the LP code. The pellet is injected at high velocity in the plasma and travels through a pedestal region. To mimic this process and to avoid exposing a pellet without any shielding cloud to an unattenuated heat flux, thus causing large numerical transients, we use the "warm-up" time concept introduced in previous works \cite{Samulyak_2007,Bosviel2020}. During the warm-up time, selected as 10 $\mu$s, the electron heat flux is linearly increased from zero to its maximum value. As our goal is to compute the steady-state ablation rate, we do not have to use a realistic pedestal crossing time: we verified that the steady-state ablation rates are not sensitive to values of the warm-up time.  During the ablation process, the pellet is surrounded by a cold, dense, neutral ablation cloud which expands isotropically near the pellet. The ablation channel is formed after about 40 $\mu$s and the flow is ionized and directed along the magnetic field lines in the longitudinal direction by the Lorentz force. The radius of the ablation channel is approximately 1.5 cm for 2 T field and becomes smaller in higher magnetic fields with stronger Lorentz force. The computational domain is a $32 cm \times 5 cm \times 5cm$ box, and the number of Lagrangian particles is on the order of $10^5$ at steady-state.
Figure \ref{lp_fixed}(a) shows the temperature distribution of the fully developed 3D pellet ablation cloud at steady state for a Neon pellet in a 2 T field and Figure \ref{lp_fixed}(b) depicts a 2D slice through the pellet center.  Figures \ref{lp_fixed} (c - e) show distributions of density, velocity, and Mach number on the same 2D slice.

\begin{figure}[H]
\centering
\includegraphics[width=0.7\columnwidth]{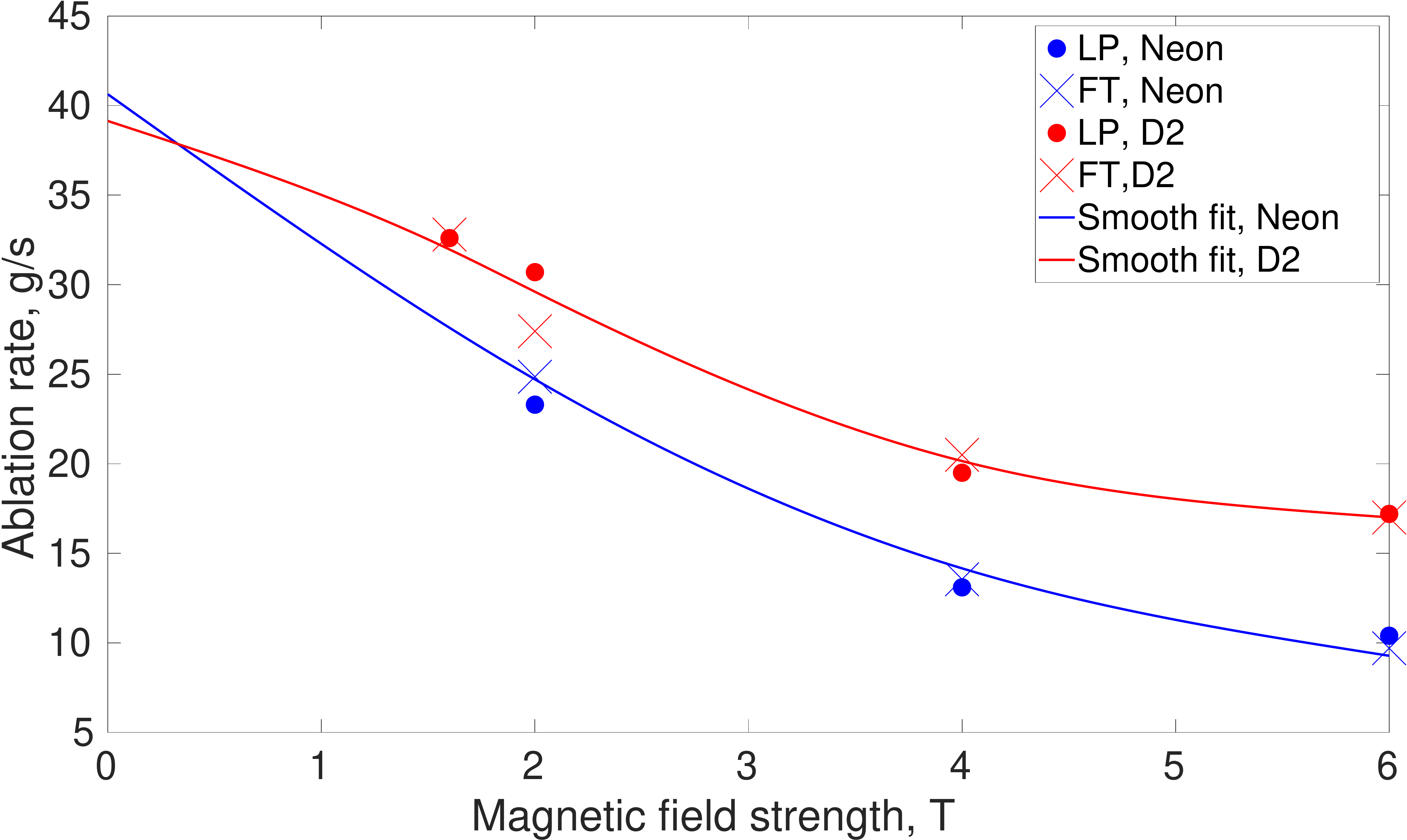}
\caption{Ablation rate as a function of \textbf{B} for neon (blue) and deuterium $D_2$ (red) pellets computed using the Lagrangian particle and  FronTier codes. }
\label{LP_FT_comparison}
\end{figure}

Tables \ref{tab:neon_fixed} and \ref{tab:d2_fixed} summarize steady state ablation rates
for the neon and deuterium 2 mm radius pellets in magnetic fields ranging from 1 to 6 Tesla in the background plasma with 2 keV temperature and $10^{14}$ 1/cc density. 
The ablation rate reduction coefficient is computed with respect to the spherically-symmetric, ideal gas EOS, semi-analytic model whose ablation rate is 
multiplied by $2/\pi$ to account for the  averaged heat flux reduction in the directional heating model compared to the spherically symmetric heat flux:
$2/\pi = \int_0^{2\pi} |\cos\alpha| d\alpha / 2\pi$, where $\alpha$ is the angle between a magnetic field line and the normal to the pellet surface.  
Comparing 3D MHD numerical data to a theoretical hydrodynamic model is convenient since such a model, that does not require  large-scale computing, is much easier to use by any researcher interested in the pellet ablation physics. To shorten explanations in the rest of the paper, we call such a reduced spherically-symmetric ablation rate as $G_{theory}^{2D}$, which 
is 41.3354 g/s for a 2 mm radius neon pellet and 39.0248 g/s for the corresponding deuterium pellet at the canonical plasma conditions.
With the fixed shielding length, the ablation rate is very sensitive to the value of the magnetic field. As the magnetic field increases, it establishes a denser and narrower ablation cloud, increasing the pellet shielding and reducing the ablation rate. 
The Lagrangian particle code is in very good agreement with the results of the FronTier pellet code (see Figure  \ref{LP_FT_comparison} and reference \cite{Bosviel2020}) on the reduction of ablation rates.
The distributions of states in the pellet cloud, shown in Figure \ref{lp_fixed}, are also in agreement with the FronTier code (see \cite{Bosviel2020} for detailed information on FronTier simulations). The deuterium pellets exhibit smaller reduction of the ablation rate  in magnetic fields of increasing strength compared to 
neon pellets. This is explained by lower densities and higher temperatures and pressures in deuterium clouds due to the absence of radiation. 
The magnetic field confines better low-pressure neon cloud that are much colder due to radiation losses. We explore more details in the next section that deals with simulations in the presence of grad-B drift.

\begin{table}[h!]
\centering
\begin{tabular}{|c|c|c|}
  \hline
  B (T)  & G (g/s) & Reduction coef. \\
  \hline
  1        &  26.24       & 1.58 \\
  2        &   23.3       & 1.77 \\
  4        &   13.1       &  3.16 \\
  5        &   11.7       &  3.53 \\
  6        &   10.4       &  3.97\\
  \hline
\end{tabular}
\caption{$Ne$ pellet ablation rates in magnetic fields of increasing strength and their reduction compared to the 2D theoretical model that predicts the ablation rate of 41.3354 g/s.}
\label{tab:neon_fixed}
\end{table}

\begin{table}[h!]
\centering
\begin{tabular}{|c|c|c|}
  \hline
  B (T)  & G (g/s) & Reduction coef. \\
  \hline
  1.6        &   32.8       & 1.19 \\
  2           &   27.4       & 1.42 \\
  4           &   20.5       &  1.90 \\
  6           &   17.0       &  2.30 \\
  \hline
\end{tabular}
\caption{Deuterium pellet ablation rates in magnetic fields of increasing strength and their reduction compared to the 2D theoretical model that predicts the ablation rate of  39.0248 g/s.}
\label{tab:d2_fixed}
\end{table}

\subsection{Simulation study of the influence of grad-B drift on the pellet ablation} \label{drift}

\begin{figure}[H]
\centering
\subfigure[Density (g/cc)]{\includegraphics[width=0.8\columnwidth]{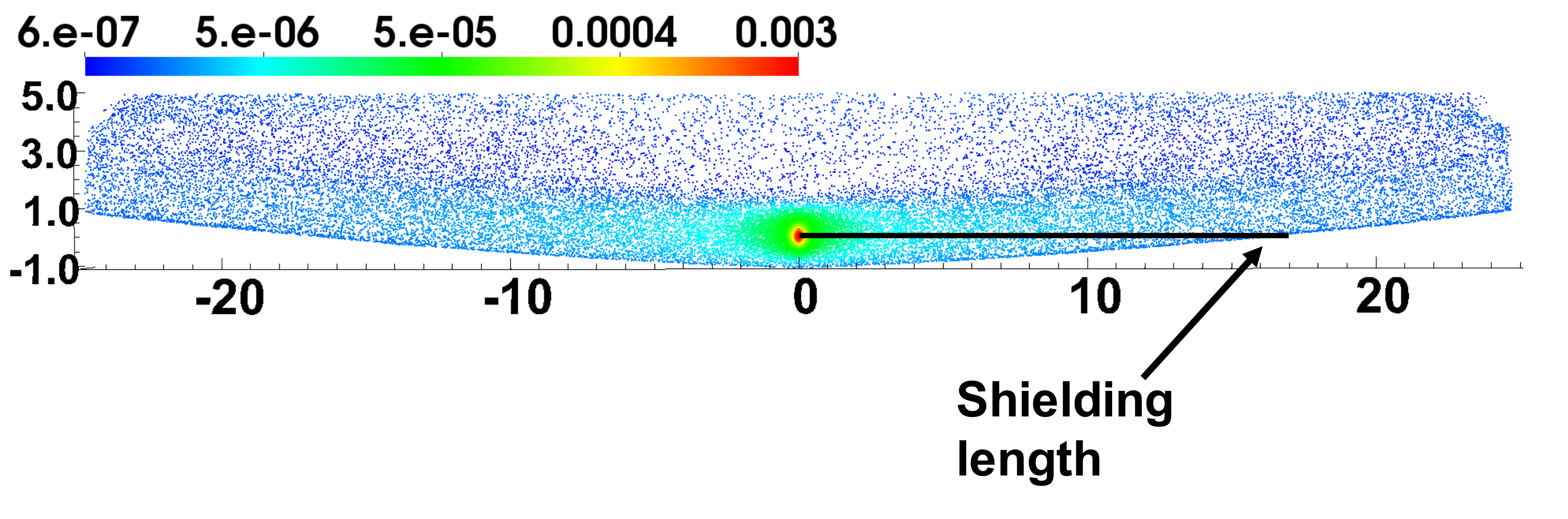}}
\subfigure[Temperature (eV)]{\includegraphics[width=0.8\columnwidth]{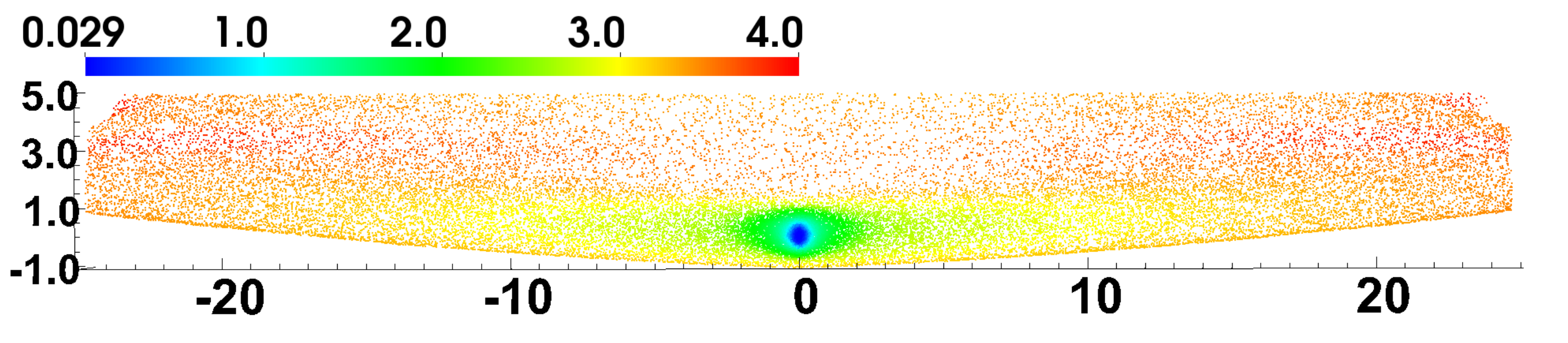}}
\subfigure[Velocity (m/s)]{\includegraphics[width=0.8\columnwidth]{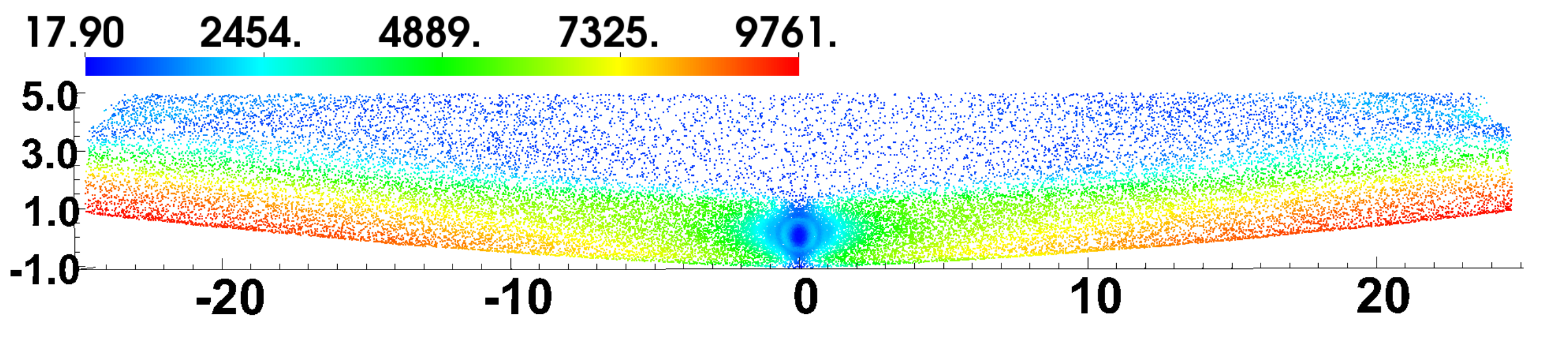}}
\subfigure[Mach number]{\includegraphics[width=0.8\columnwidth]{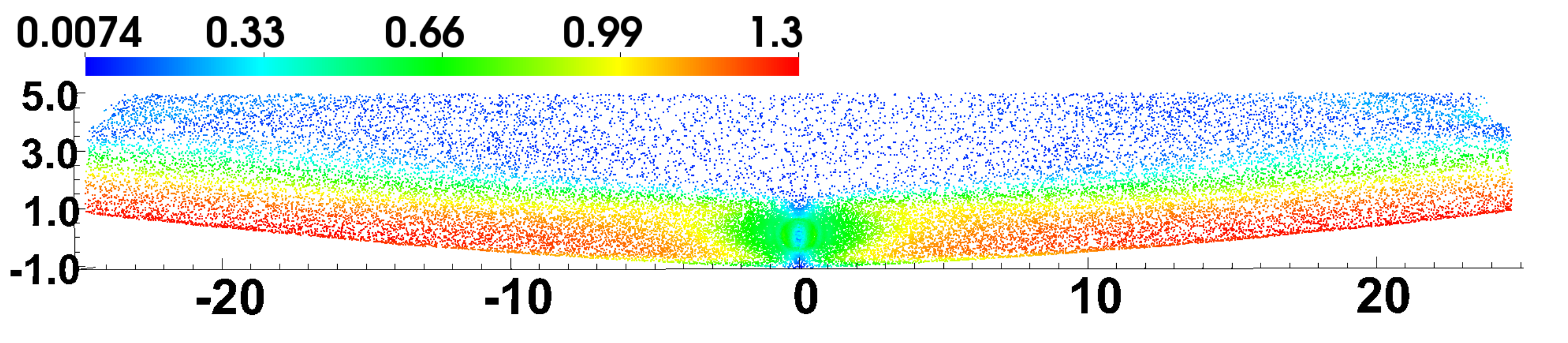}}
\caption{Distribution of ablation cloud density (a), temperature (b), velocity (c), and Mach number (d) at steady state  for neon pellet in B = 2 T field with grad-B drift. Horizontal axes (cm) are along the magnetic field, and vertical axes (cm) are transverse to the magnetic field in the grad-B drift direction.}
\label{fig:neon_drift}
\end{figure}

Having verified the Lagrangian particle pellet code against the theoretical semi-analytic model and the FronTier code, we use it now for more realistic 3D simulations in which the 
pellet cloud shielding length is established self-consistently via the grad-B drift. This process is illustrated in Figure \ref{fig:neon_drift} which depicts a 2D slice of density, temperature, velocity and Mach 
number for a 2 mm radius neon pellet in 2 T magnetic field with the DIII-D major radius of 1.6 m and the canonical plasma parameters.

\begin{figure}
\centering
\subfigure[Neon pellet]{\includegraphics[width=0.7\columnwidth]{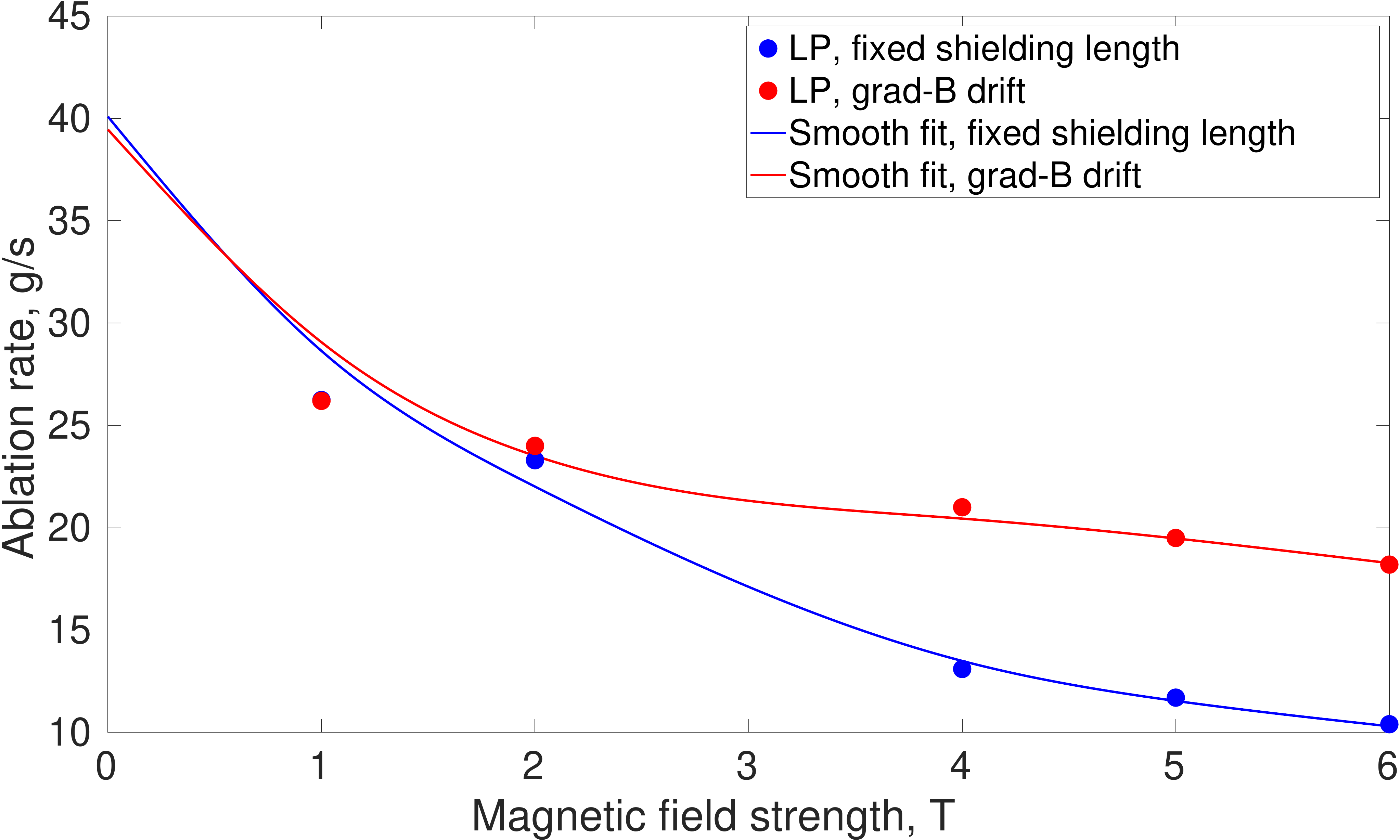}}
\subfigure[Deuterium pellet]{\includegraphics[width=0.7\columnwidth]{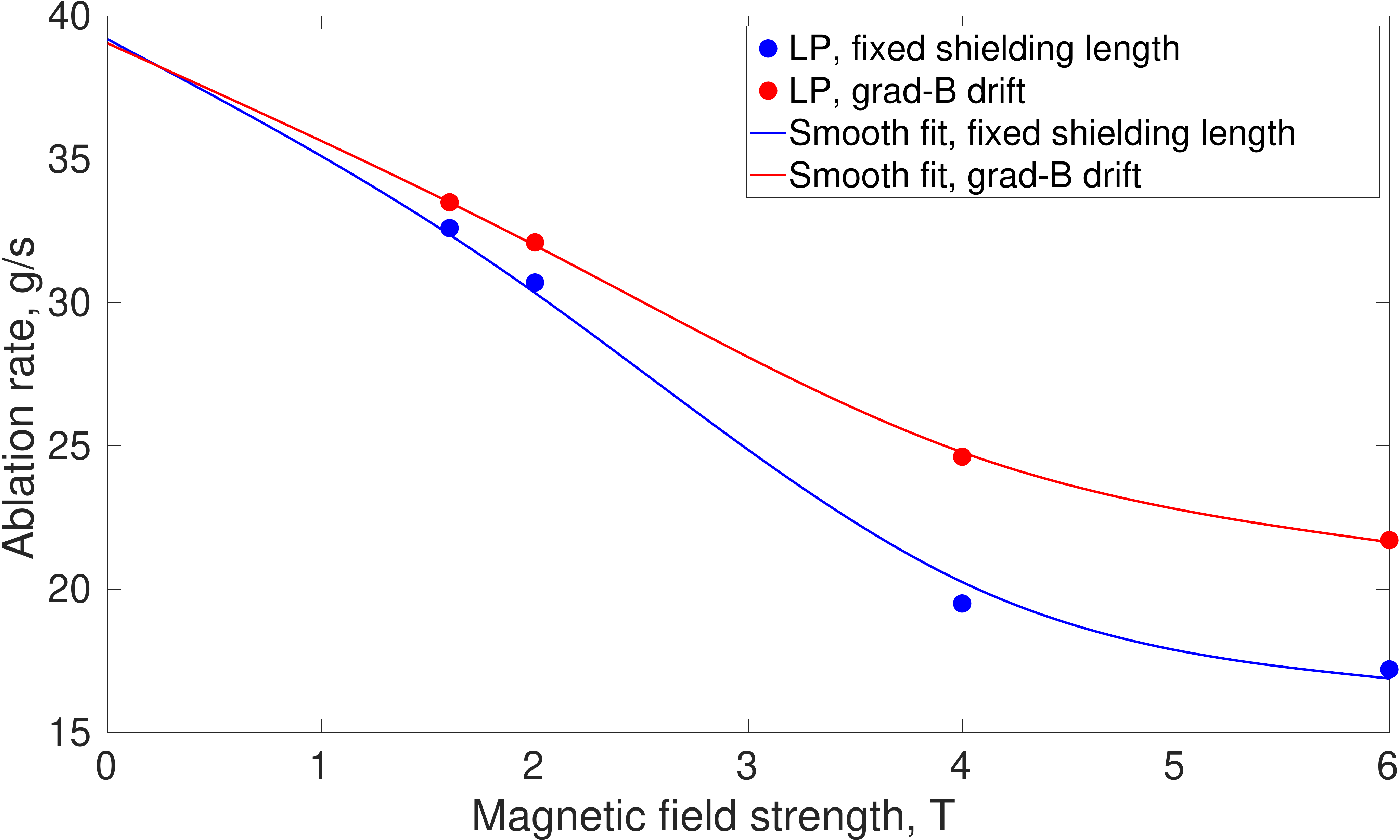}}
\caption{Comparison of ablation rates in the magnetic field for fixed shielding length simulations and grad-B drift simulations for (a) neon pellet, and (b) deuterium pellet. Data points from simulations with fixed shielding length are shown with blue dots, simulation data points with grad-B drift are shown with red dots, and solid lines represent polynomial fits.}
\label{fig:G_drift1}
\end{figure}

Our results demonstrate that the grad-B drift has a major influence on the pellet ablation in magnetic fields. Main observations are summarized in tables that provide pellet ablation rates and shielding lengths for neon (Table \ref{table_drift_neon}) and  deuterium  (Table \ref{table_drift_d2}) pellets using the DIII-D major radius of 1.6 m, and Figure 
\ref{fig:G_drift1} that compares these ablation rates with ones obtained using the fixed shielding length approximation in cylindrically-symmetric simulations. As before, the reduction coefficient was computed using $G^{2D}_{theory}$.

\begin{table}[ht]
\centering
\begin{tabular}{|c|c|c|c|}
\hline
B, T & Shielding length, cm & Ablation rate, g/s & Reduction coefficient \\
\hline
1  & 21 & 26.2 & 1.58 \\
\hline
2  & 17 & 24  &  1.72 \\
\hline
4  & 15 & 21  & 1.97 \\
\hline
5  & 13.5 & 19.5 &  2.12 \\
\hline
6  & 12.5 & 18.2 & 2.27 \\
\hline\end{tabular}
\caption{Shielding lengths and ablation rates in grad-B drift LP simulations of neon pellet in various magnetic fields with the DIII-D tokamak major radius.}
\label{table_drift_neon}
\end{table}

\begin{table}[ht]
\centering
\begin{tabular}{|c|c|c|c|}
\hline
B, T & Shielding length, cm & Ablation rate, g/s & Reduction coefficient \\
\hline
1.6  & 25 & 33.5 & 1.16 \\
\hline
2  & 23 & 32.1 & 1.21\\
\hline
4  & 17 & 24.62 &1.59\\
\hline
6 & 13.5 & 21.71 & 1.80\\
\hline\end{tabular}
\caption{Shielding lengths and ablation rates in grad-B drift LP simulations of deuterium ($D_2$) pellet in various magnetic fields with the DIII-D tokamak major radius.}
\label{table_drift_d2}
\end{table}
\vspace{0.15in}

Figure \ref{fig:G_drift1}(a)  shows that the reduction of the neon pellet ablation rate in magnetic fields of increasing strength is much weaker compared to simulations with the fixed shielding length. The same tendency is also exhibited by the deuterium pellet \ref{fig:G_drift1}(a), but to a less extent. For the neon pellet (see Table \ref{table_drift_neon}), the coefficient of ablation rate reduction with respect to $G^{2D}_{theory}$ is about the same for 2 Tesla field: 1.72 compared to  
1.77 for the fixed shielding length. But with the increase of the magnetic field to 6 Tesla, the reduction coefficient is only 2.27 while it was 3.97 for the fixed shielding length. 
For deuterium (see Table \ref{table_drift_d2}), this change of the ablation rate reduction is less significant: 1.16 reduction coefficient at 1.6 Tesla is close to 1.19 for the fixed shielding length while these quantities for 6 Tesla field become 1.80 and 2.30, correspondingly. 

\begin{figure}[H]
\centering
\subfigure[Fixed shielding length]{\includegraphics[width=0.7\columnwidth]{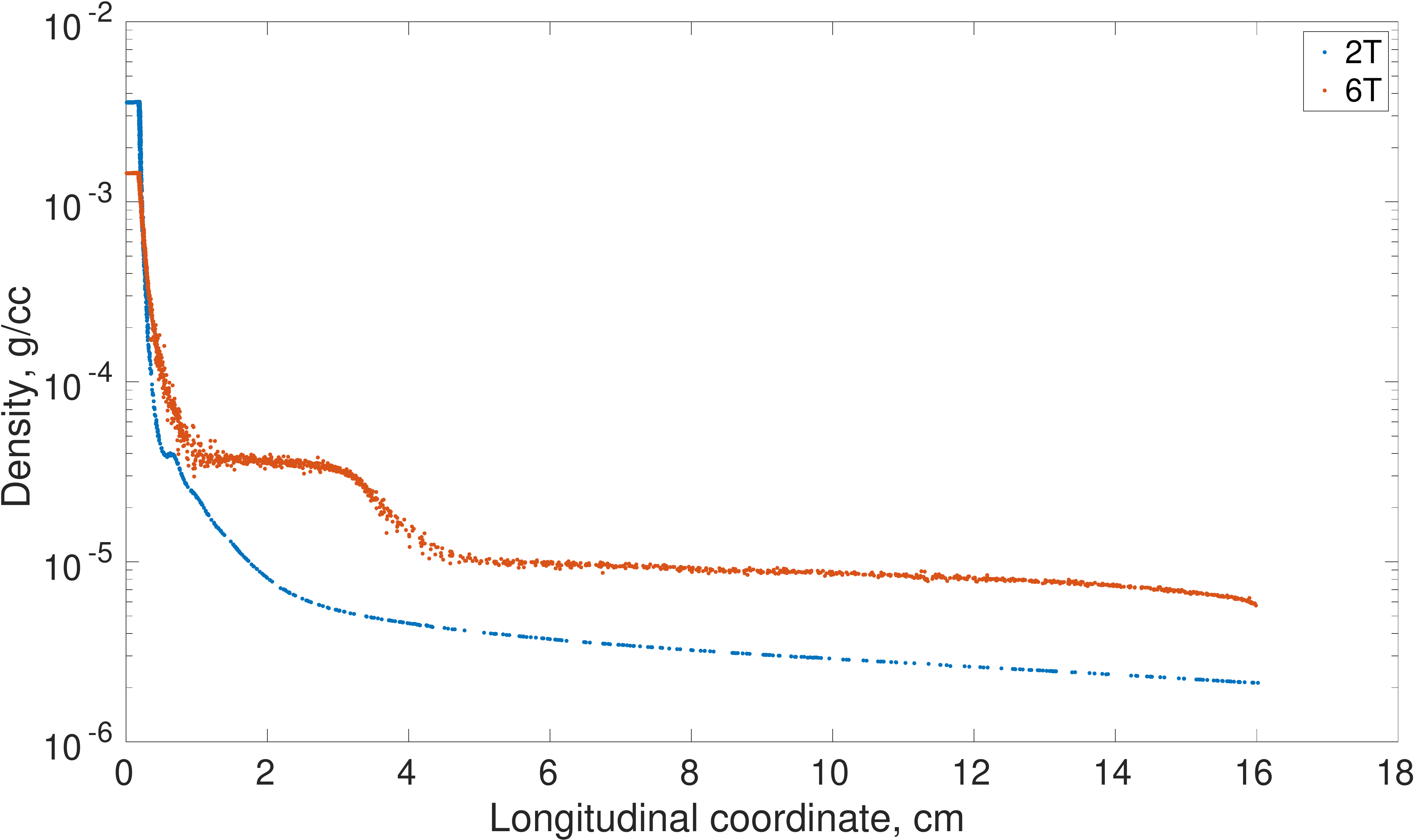}}
\subfigure[Shielding length computed from grad-B drift]{\includegraphics[width=0.7\columnwidth]{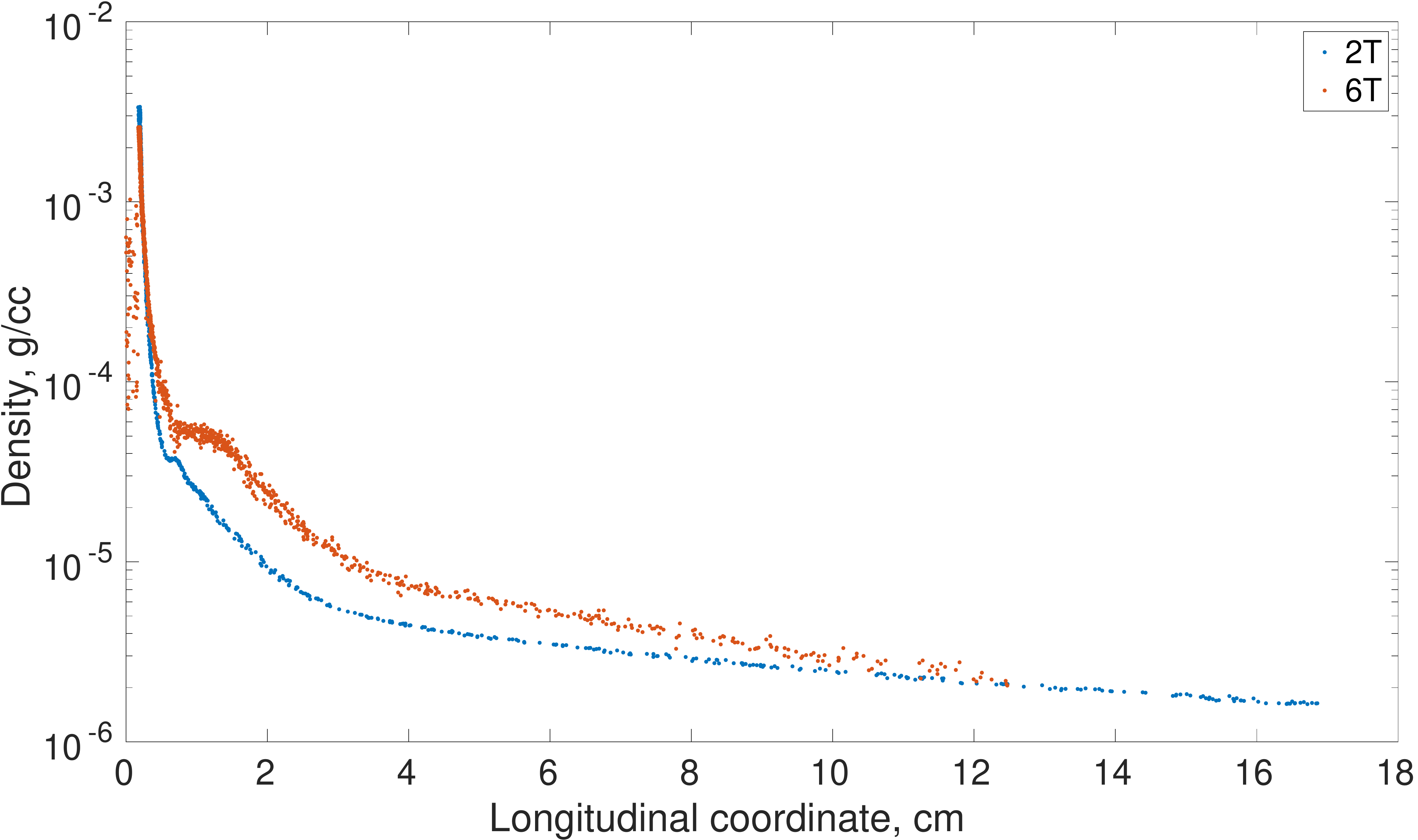}}
\caption{Comparison of ablation cloud density along magnetic field line passing through the pellet center for 2T and 6T magnetic fields for (a) simulations with fixed shielding length, and (b) shielding length established by grad-B drift. Blue points plot simulation data with 2T field and red points plot simulation data with 6T field.}
\label{fig:dens_2v6}
\end{figure}

The 1st reason for a much weaker effect of the magnetic field-induced reduction of the ablation rate is clear from the shielding length column in tables \ref{table_drift_neon} and \ref{table_drift_d2}:
the shielding length decreases with the  increase of the magnetic field, reducing the pellet shielding and increasing the amount of energy that reaches the pellet surface. While this is an important factor, it is not the most significant: density of the ablated material is low at the end of the shielding  column, and the change of the shielding length can not fully account for very different ablation rates of the neon pellets at 6 Tesla with and without grad-B drift. The main reason for this effect is in the change of overall properties of the ablation cloud. Without the grad-B drift, the ablated material is completely confined to a narrow ablation channel by the Lorentz force. With the increase of the magnetic field, the ablation channel narrows, increasing the density in the channel, the pellet shielding, and reducing the ablation rate. Such a confinement does not take place in the presence of grad-B drift : the ablated material drifts across magnetic field lines and makes the shielding effect less sensitive to the magnetic field change. This statement is convincingly demonstrated in Figure \ref{fig:dens_2v6} that plots a comparison of the ablation cloud density along the magnetic field line passing through the pellet center  for 2T and 6T magnetic fields for (a) simulations with fixed shielding length, and (b) shielding length established by grad-B drift. The density profile for 2 T field is approximately the same for both simulations. Indeed, the ablation rates are very close (23.3 g/s for the fixed shielding length and 24 g/s for the simulation with grad-B drift) as are the shielding lengths: the grad-B drift simulation establishes 17 cm shielding length while 16 cm was assumed for fixed shielding length simulations. But with the increase of the magnetic field to 6 T, the density is significantly higher in the case of the fixed shielding length.  The horizontal plateau in Figure \ref{fig:dens_2v6}(a)  with value  
$\sim 4\times 10^{-5}$ g/cc between 1 and 3.5 cm represents the region where intense ionization takes place. This ionized material, which still maintains higher pressure, partially drifts by the grad-B force and the density quickly falls down (see Figure \ref{fig:dens_2v6}(b)). The almost constant density in the far-field (5 - 16 cm) of the fixed-length case corresponds to the  decreasing density 
in plot (b) where the drift force is increased by the Mach number component of (\ref{eq:drift}).

\begin{figure}[H]
\centering
\subfigure[Density (g/cc)]{\includegraphics[width=0.49\columnwidth]{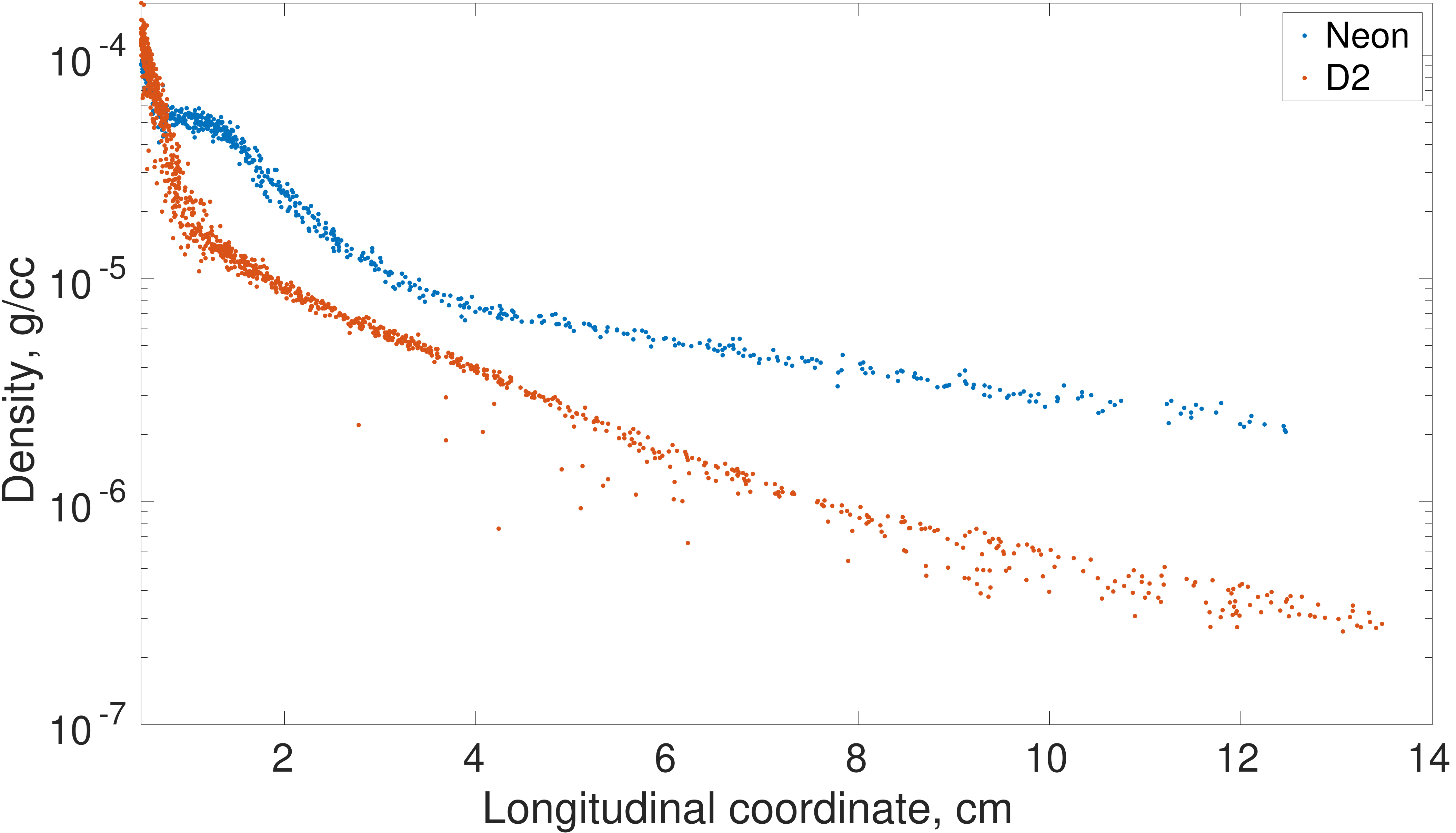}}
\subfigure[Pressure (bar)]{\includegraphics[width=0.49\columnwidth]{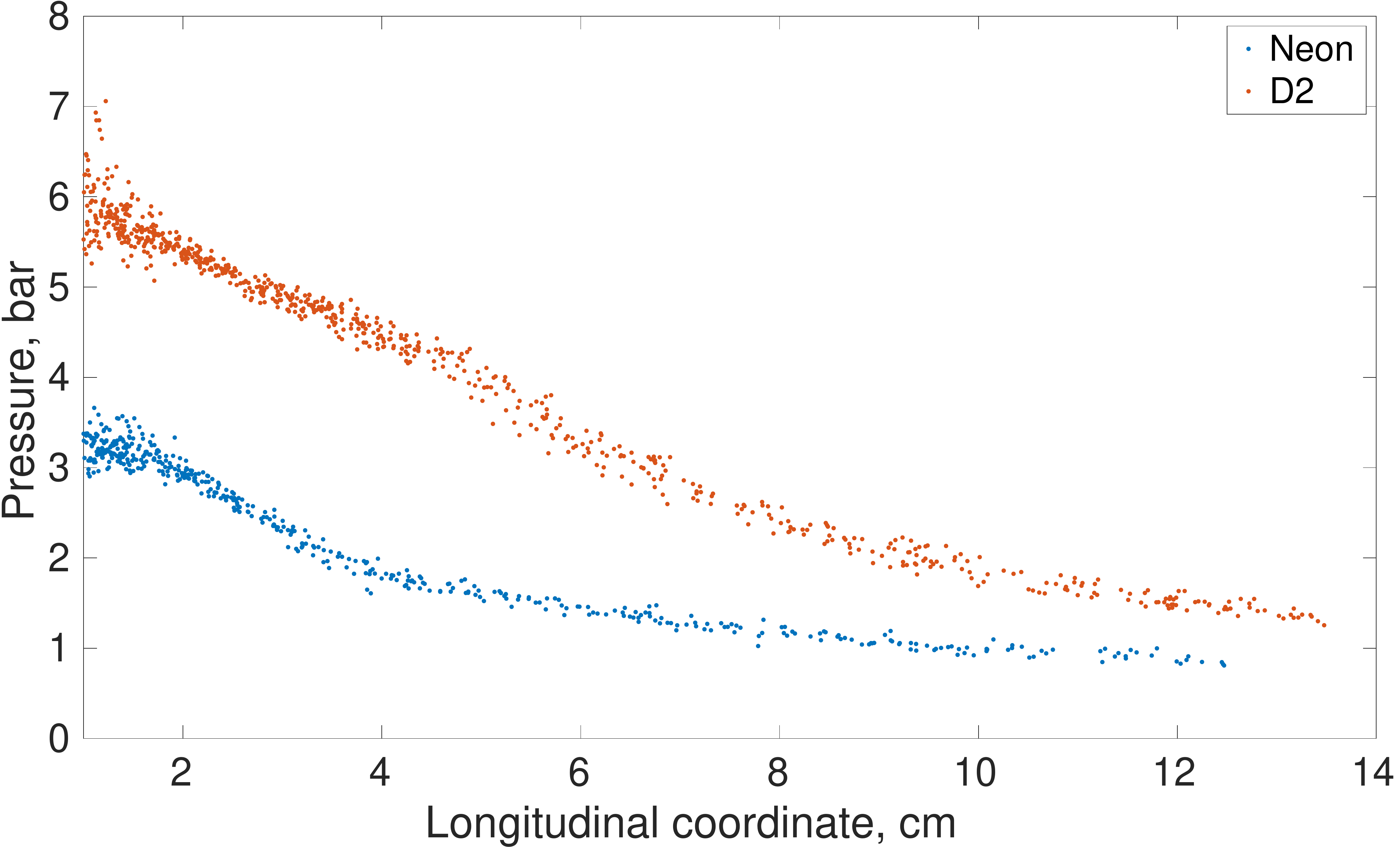}}
\subfigure[Mach number]{\includegraphics[width=0.49\columnwidth]{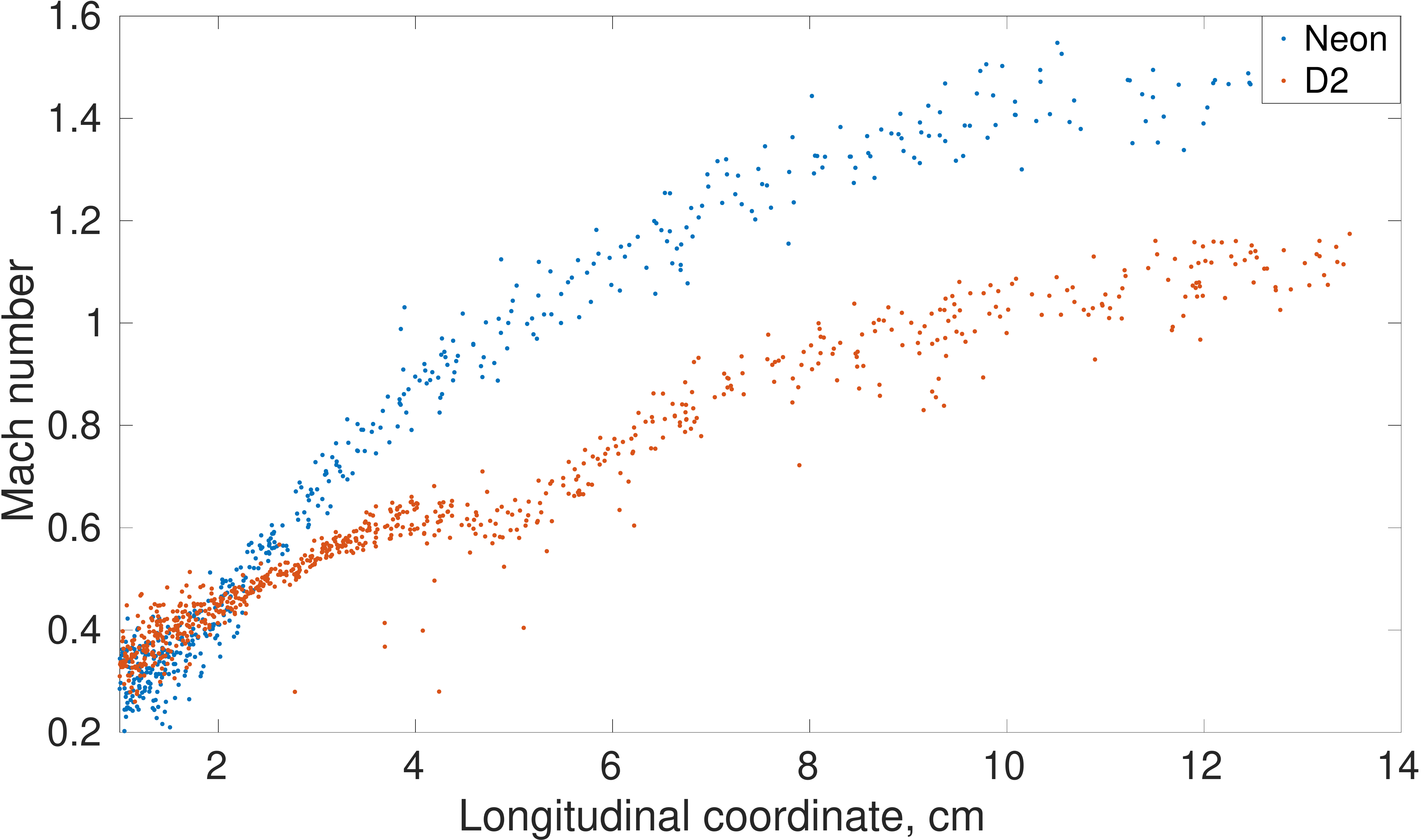}}
\subfigure[Longitudinal velocity (m/s)]{\includegraphics[width=0.49\columnwidth]{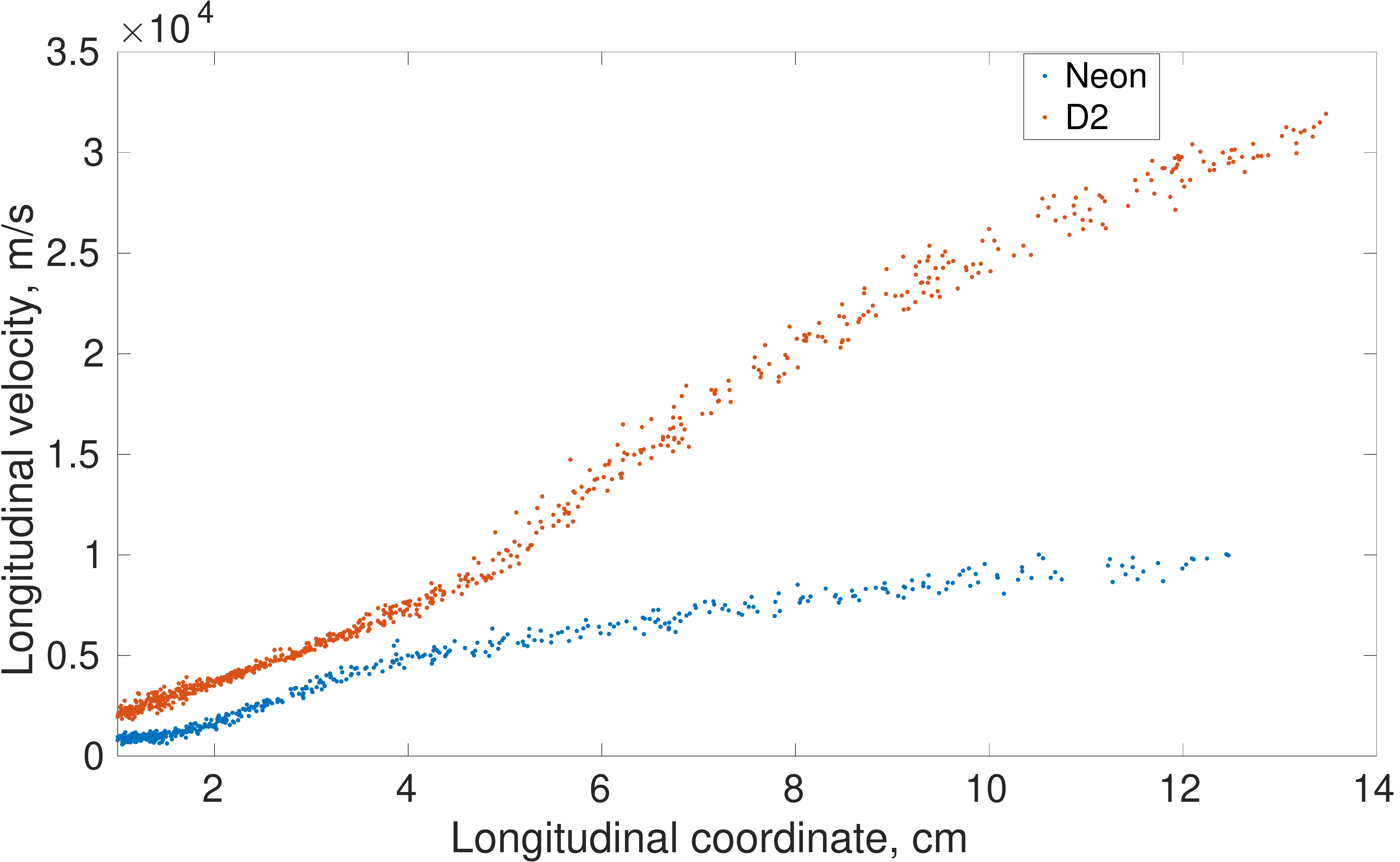}}
\caption{Comparison of neon and deuterium cloud states in 6 T field along the magnetic field line passing through the pellet center. Blue points plot neon simulation data and red points plot deuterium simulation data}
\label{fig:neon_d2_drift}
\end{figure}

\begin{figure}[H]
\centering
\subfigure[Neon cloud drift velocity]{\includegraphics[width=0.8\columnwidth]{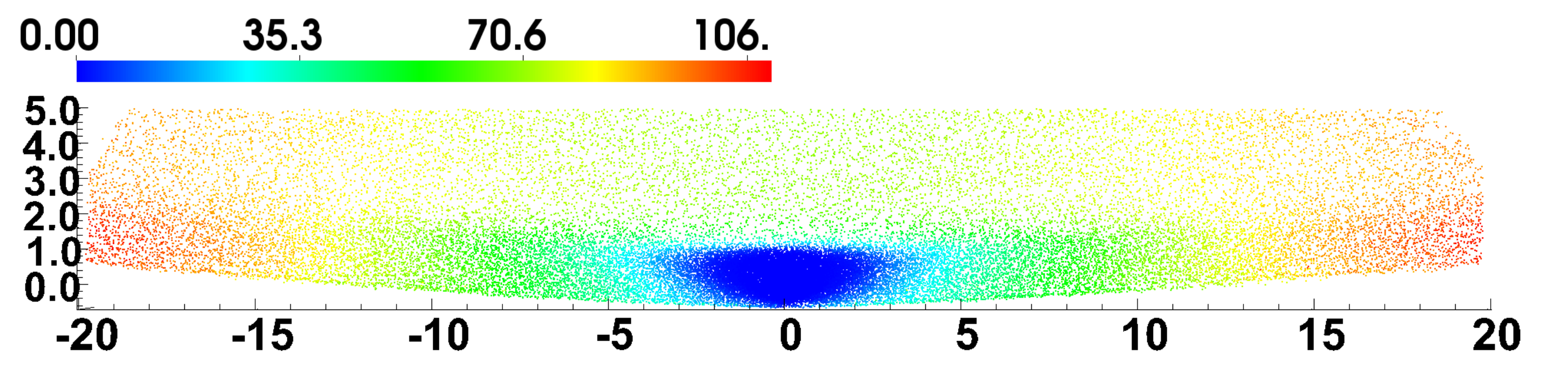}}
\subfigure[Deuterium cloud drift velocity]{\includegraphics[width=0.8\columnwidth]{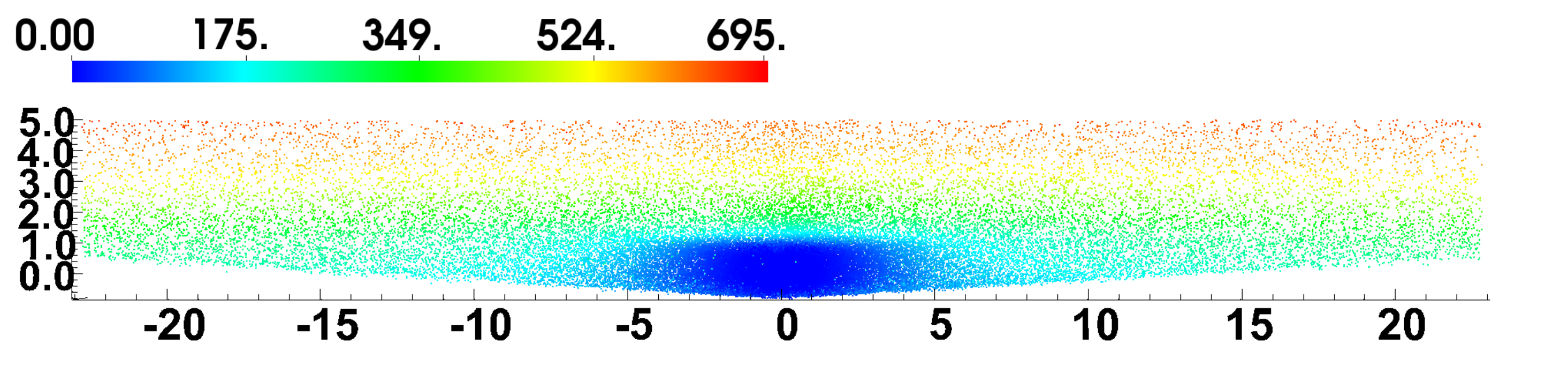}}
\caption{Two-dimensional distribution of drift velocity in the neon ablation cloud (a) and deuterium cloud (b). Horizontal axes (cm) are along the magnetic 
field, and vertical axes (cm) are transverse to the magnetic field in the grad-B drift direction.}
\label{fig:drift_v_2d}
\end{figure}

It is interesting to note that despite totally different properties of the neon and deuterium pellet ablation clouds,  they exhibit relatively similar values of the shielding length which seem to converge at higher magnetic fields: the shielding length for neon in 5 T field is the same as the deuterium shielding length at 6 T, and their difference is only 1 cm at 6 T. To provide some explanations of this result, we compare thermodynamic states of both clouds in 6T field along the magnetic field line passing through the pellet center in Figure \ref{fig:neon_d2_drift}. The deuterium cloud density is obviously much lower, contributing to a higher drift acceleration. Pressure in the deuterium cloud is higher compared to the neon cloud (the absence of radiation increases temperature),  which also increases the grad-B drift (\ref{eq:drift}). While the Mach number factor is higher for the neon cloud due to radiation cooling and lower temperatures in neon, this effect only slightly affects the overall result. As a consequence, the drift velocity of the deuterium cloud is much higher compared to the neon cloud, as shown in
Figure \ref{fig:drift_v_2d}. But in combination with much higher longitudinal velocity of deuterium (Figure \ref{fig:neon_d2_drift}(d)),
trajectories of deuterium particles are not much different compared to neon particle trajectories, leading to similar shielding lengths.

\begin{figure}[H]
\centering
\includegraphics[width=1.0\columnwidth]{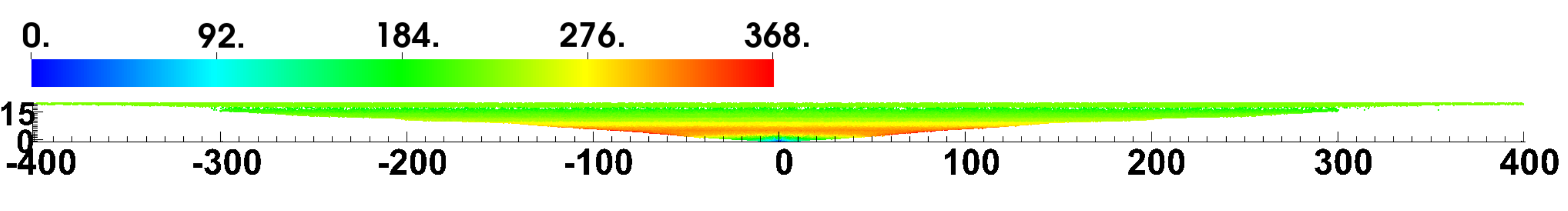}
\caption{Two-dimensional distribution of drift velocity in deuterium cloud at large spatial scales. Horizontal axis (cm) is along the magnetic field, 
and the vertical axis (cm) is transverse to the magnetic field in the grad-B drift direction.}
\label{fig:long_d2_cloud}
\end{figure}

We would like to comment now on the influence various terms in the grad-B drift expression (\ref{eq:drift}) on the drift velocity and the pellet ablation process. The pressure difference was recognized as the main driving force in earlier works \cite{Parks_2000,Rozhanskij_1994,Pegourie_HPI2}. The Mach number contribution, introduced in
\cite{Parks_2005}, is very important for neon pellets as the radiation cooling reduces the temperature and the sound speed and increases the Mach number. This term is less important but not negligible for deuterium pellets. Our results show that Alfven wave drag is negligibly small for both neon and deuterium pellets in simulations of ablation clouds in close proximity to the pellet in the direction transverse to the magnetic field. Only domain sizes that slightly exceed the shielding length in the longitudinal direction and extend several centimeters beyond the pellet in the transverse direction, as shown in Figure \ref{fig:neon_drift}, are relevant for the computing of steady-state pellet ablation rates. If, however, one is interested in parallel and transverse expansions of the ablation flow at large length scales, the Alfven wave drag becomes an important factor. 
In Figure \ref{fig:long_d2_cloud}, we show long-scale dynamics of the ablation cloud of a 2 mm radius deuterium pellet in the background plasma with 500 eV temperature,
$4\times 10^{13}$ 1/cc density, and 2 Tesla magnetic field. The drift velocity shown in the figure was strongly limited by the Alfven wave drag: without the Alfven wave drag term, the outer layers develop unrealistically large drift velocities. 

\begin{figure}[H]
\centering
\subfigure[Ablation rates]{\includegraphics[width=0.7\columnwidth]{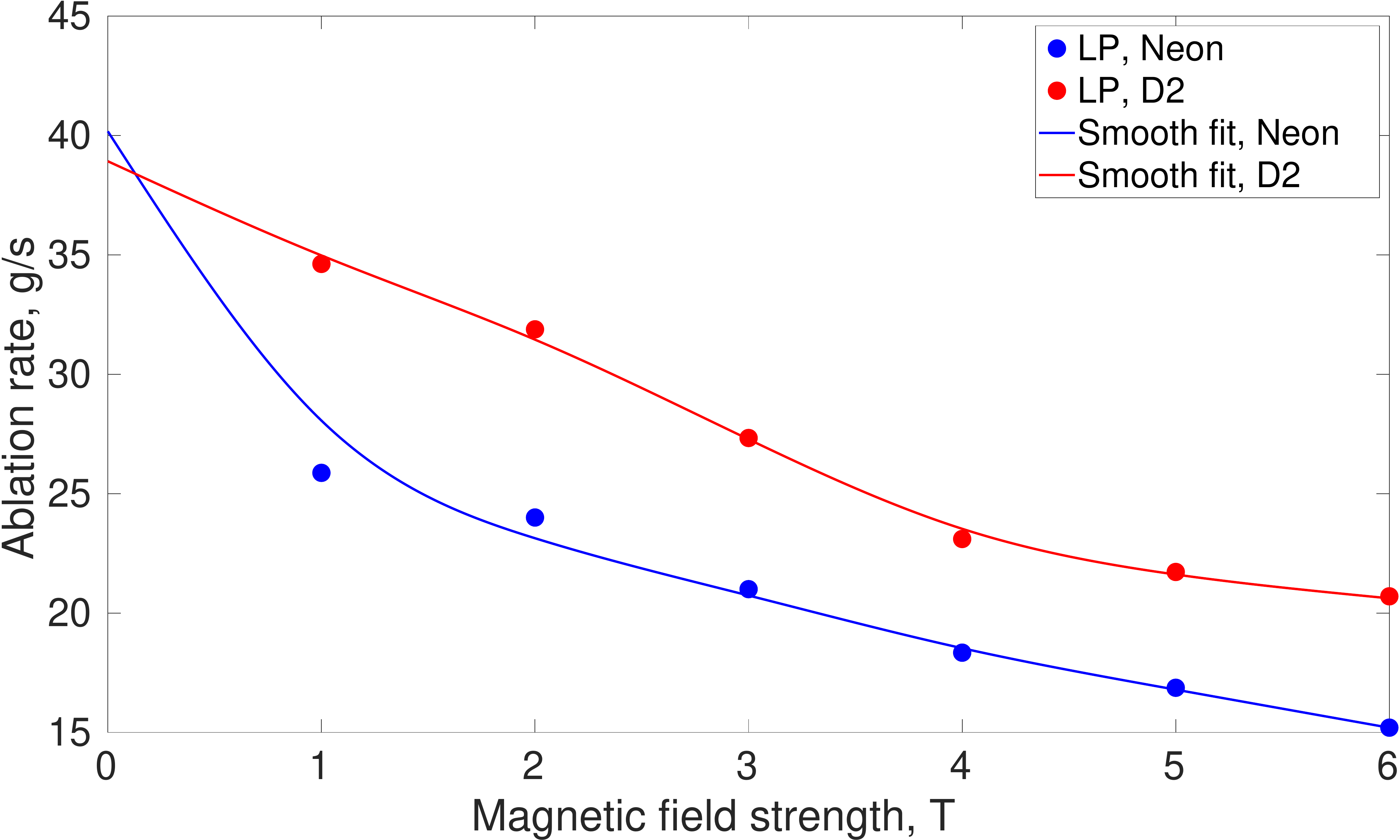}}
\subfigure[Shielding lengths]{\includegraphics[width=0.7\columnwidth]{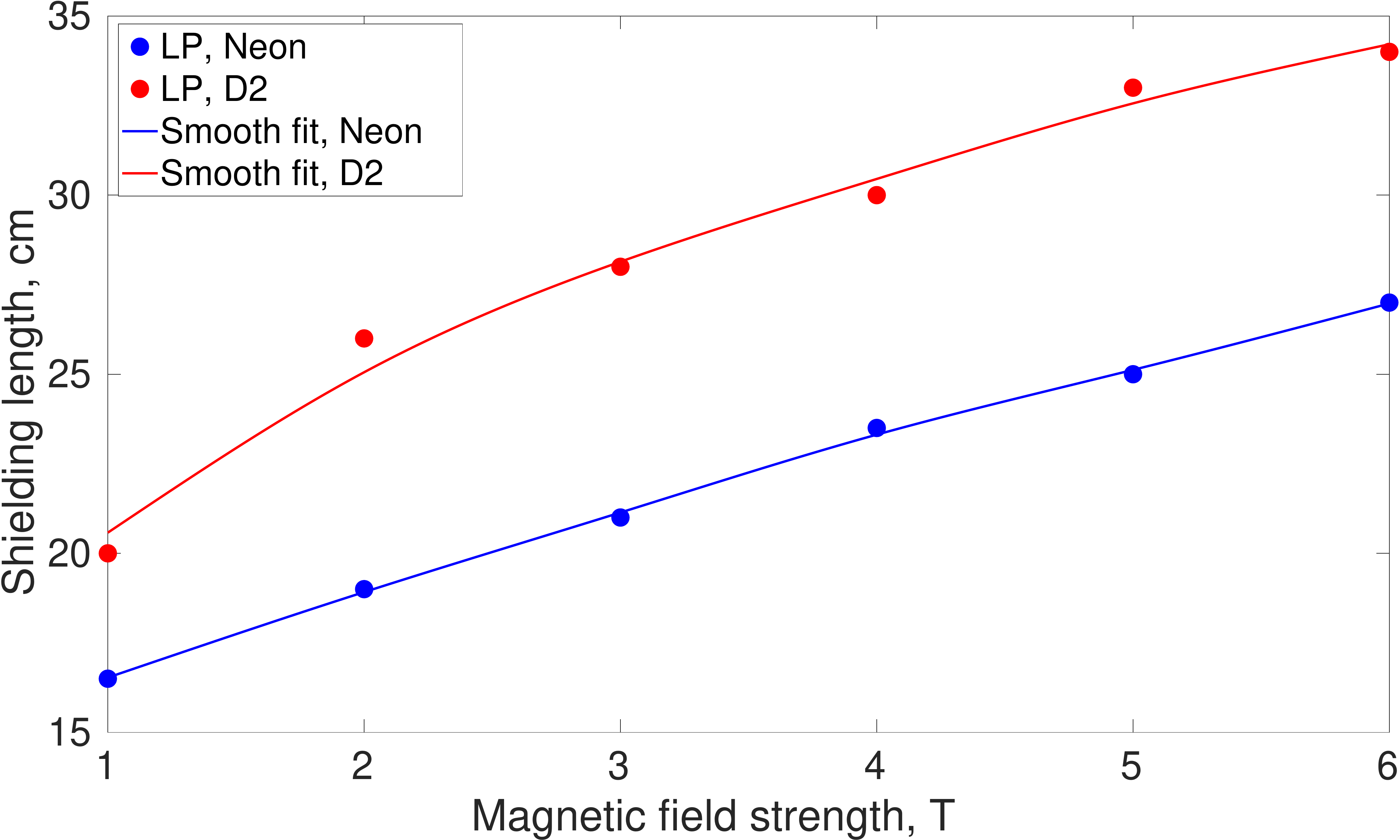}}
\caption{Ablation rates (a) and shielding lengths (b) for 2 mm neon and deuterium pellets in magnetic fields of increasing strength at constant ratio B[T]/R[m]=1.}
\label{fig:G_RB1}
\end{figure}

In this section, we demonstrated that both the magnetic field strength and the grad-B drift have a strong influence on the pellet ablation rate. The most important tokamak parameter influencing the grad-B drift is the major radius $R$. We notice that the ratio of the peak magnetic field in Tesla to the tokamak major radius in meters is close to unity for all practical tokamaks, including ITER. In order to reduce dimensionality of the parameter space for simulations of practical interest, we performed simulations using magnetic fields in the range from 1 to 6 Tesla by imposing the constraint $R[m] = B[Tesla]$. Results for 2 mm radius neon and deuterium pellets are shown in Figure \ref{fig:G_RB1}. 
A simulation database for neon pellet ablation rates spanning a range of plasma densities and temperatures and pellet radii will be computed in the future.

\subsection{Simulation of SPI}

All numerical models implemented in the Lagrangian particle code are designed to handle the ablation of multiple SPI fragments. While the physics part of this paper focuses on detailed simulation studies of the ablation rate of single pellets in the presence of magnetic field and grad-B drift, we would like to demonstrate our capabilities of fully-resolved SPI simulations. 

\begin{figure}[H]
\centering
\subfigure[Velocity distribution (x10, m/s)]{\includegraphics[width=0.8\columnwidth]{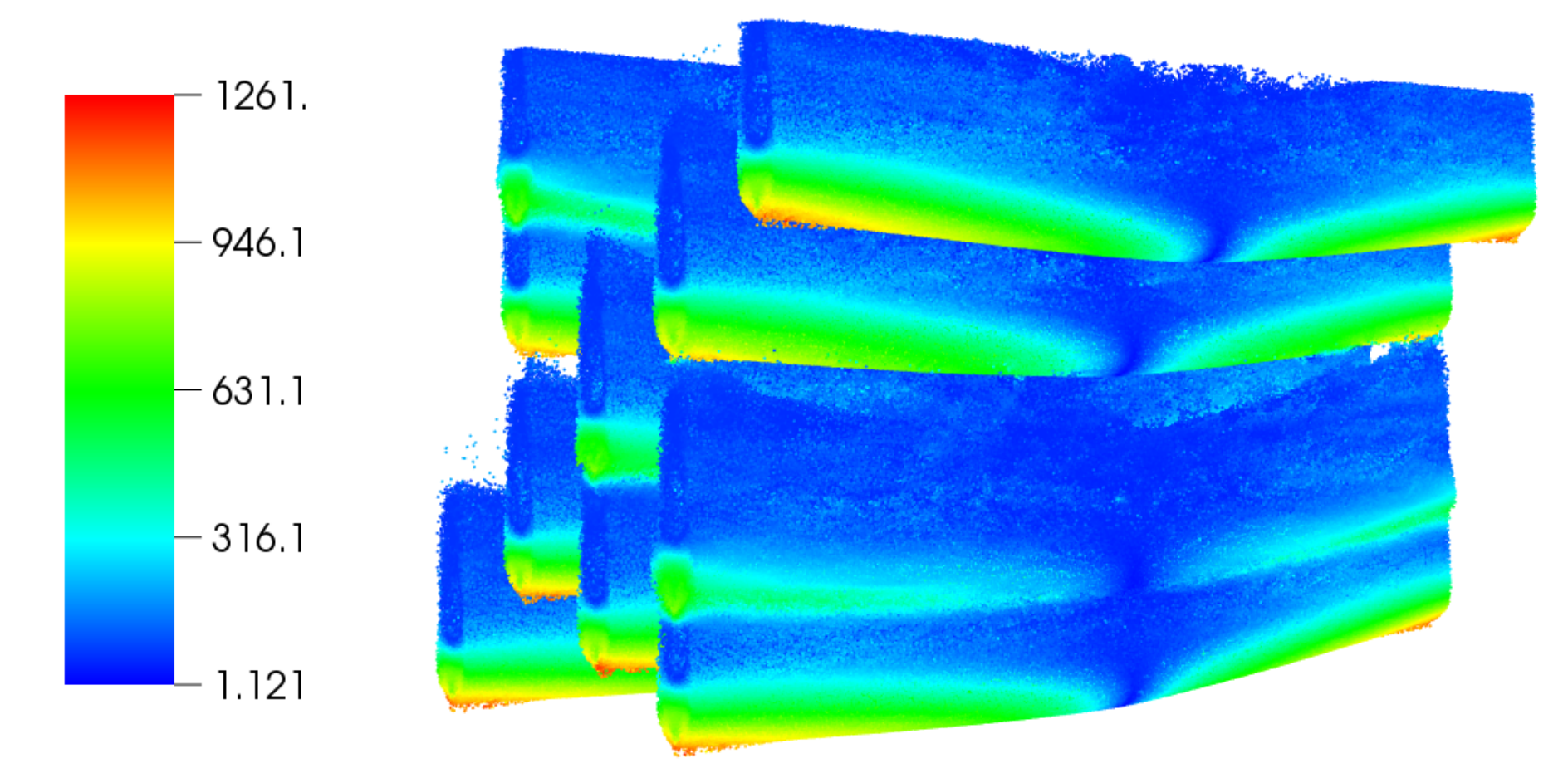}}
\subfigure[Quadtree for finding integration lines]{\includegraphics[width=0.8\columnwidth]{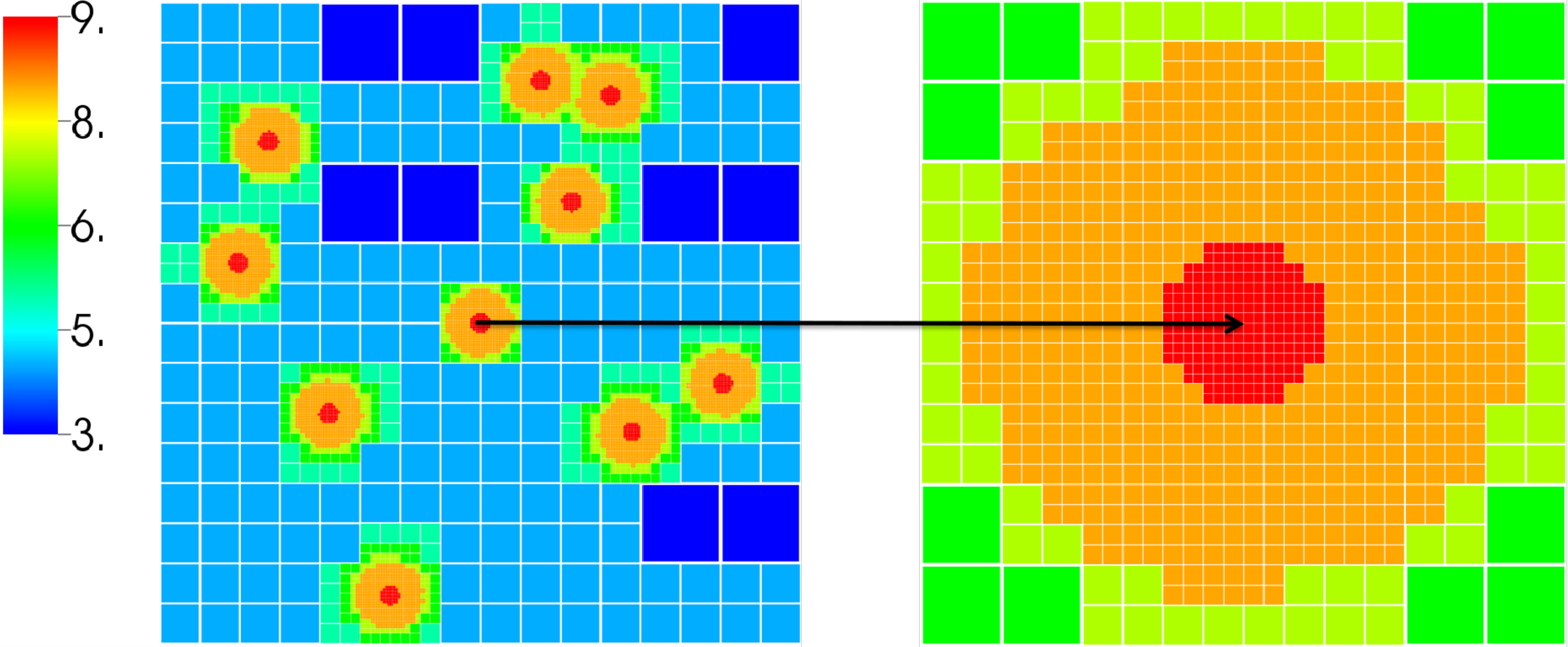}}
\caption{(a) SPI simulation involving 10 neon fragments with MHD forces and grad-B drift.  Velocity distribution is shown. (b) Quadtree for finding integration lines along magnetic field: global view is shown on the left and detailed structure around the center of an SPI fragment is shown on the right. The colormap provides quadtree levels. Each cell shown in the figure (including most refined cells shown in red) contains one  integration line used in the kinetic heating and gradB drift models.}
\label{fig:spi}
\end{figure}

We estimate that in an SPI injection experiment into DIII-D, a large pellet with the total neon inventory of 0.0213 moles was shattered into approximately 250 fragments with the average size of 0.66 mm, and the average distance between fragments was of the order of 4 - 5 cm. While the Lagrangian particle code is capable of resolving a large number of fragments, it is also possible to simulate a subset of the whole system that includes 10 - 50  interacting fragments and use periodic or other suitable boundary conditions. In the example below, we simulate ablation of 10 fragments with uniform radius of 0.66 mm randomly distributed in a computational box of the corresponding volume. We observe that fragments that partially screen each other either directly or through the grad-B drift of the ablated material interact and affect the ablation rate. 
Figure \ref{fig:spi}(a) depicts a 3D distribution of the velocity field in the ablating plume. Figure \ref{fig:spi}(b) illustrates the critical component of the SPI simulation: the adaptive selection of integration lines for the kinetic heating and grad-B drift models by constructing a quadtree for all particles in a plane transverse to the magnetic field.  Each cell shown in the figure (including most refined cells shown in red) contains one  integration line. This example is intended only for the demonstration of computational capabilities of the code. Simulations covering a range of practically important parameters and detailed analysis of physics results will be the subject of a forthcoming paper.
 
\section{Conclusions}

3D numerical model for the ablation of pellets and shattered pellet injection (SPI) fragments in tokamaks in the plasma disruption mitigation and fueling parameter space has been developed based on the Lagrangian Particle (LP) code \cite{SamWangChen2018}.
The code implements the low magnetic Reynolds number MHD equations, kinetic models for the electronic heating, a pellet surface ablation model, equation of state with multiple ionization support, radiation and a model for grad-B drift of the ablated material across the magnetic field. The Lagrangian particle algorithm is highly adaptive, capable of simulating a large number of fragments in 3D while eliminating numerical difficulties of dealing with the tokamak background plasma. The Lagrangian approach makes it much easier to extract relevant data for a multiscale coupling with tokamak-scale MHD codes, details of which will be presented in a forthcoming paper. The code has achieved good agreement with the semi-analytic model for spherically symmetric ablation flows. Axisymmetric simulations of neon and deuterium pellets in magnetic fields ranging from 1 to 6 Tesla have been compared with previous simulations using FronTier, a grid-based Eulerian code with
explicit tracking of material interfaces, and good agreement has also been obtained. Axisymmetric simulations require an additional parameter, the shielding length, to limit the total length of the cloud and establish steady-state ablation rates. Simulations with a fixed shielding length show a strong reduction
of both neon and deuterium pellet ablation rates in magnetic fields of increasing strengths. With the increase of the magnetic field, the ablation channel narrows, the density and the pellet shielding increase, and the ablation rate reduces. 

The main physics contribution of the paper is a detailed study of the influence of 3D effects, in particular the grad-B drift, on pellet ablation rates and properties of ablation clouds. Smaller reductions of ablation rates in magnetic fields compared to axially symmetric simulations have been observed because the ablated material is not confined to narrowing channels in the presence of grad-B drift. Another factor contributing to smaller reduction of the ablation rate is the magnetic field dependence of the 
shielding length which decreases with the increase of the magnetic field. Since the ratio of the tokamak peak magnetic field in Tesla to the major radium in meters is close to unity for practical machines, including ITER, simulations in magnetic fields ranging from 1 to 6 Tesla have been performed subject to the constraint B[T]/R[m] = 1. A
simulation pellet ablation database will be compiled in the future. 
  
Contributions of various factors in the grad-B drift model have been quantified. It was shown that  the effect of the centrifugal force arising from the parallel flow velocity with the Mach number $M_\parallel$ in the curved toroidal magnetic field is very important for the grad-B drift of neon pellet ablation clouds. It is less important for deuterium clouds as the absence of radiation increases temperature and reduces the Mach number. The Alfven wave drag has a very small effect on pellet ablation rates, but it significantly 
influences long-scale propagation of the ablated material. 

We concluded this paper with an example of SPI simulation involving 10 fragments subject to periodic boundary conditions. This example was intended only for the demonstration of computational capabilities of the code. Simulations covering a range of practically important parameters and detailed analysis of physics results will be the subject of a forthcoming paper.

Future work will focus on simulations of composite deuterium - neon pellets, ablation of pellets affected by runaway electrons, fully resolved SPI simulations, studies of ablation rate scaling laws in magnetic fields, and multiscale coupling  with tokamak-scale MHD codes.

\vskip5mm
{\bf Acknowledgement.}
This research has been supported by the Center for Tokamak Transient Simulations with the US DOE  SciDAC program.

\vskip5mm

\bibliographystyle{ieeetr}
\bibliography{references}

\begin{thebibliography}{10}

\bibitem{Commaux16}
N.~Commaux, D.~Shiraki, L.~Baylor, E.~Hollmann, N.~Eidietis, C.~Lasnier,
  R.~Moyer, T.~Jernigan, S.~Meitner, S.~Combs, and C.~Foust, ``First
  demonstration of rapid shutdown using neon shattered pellet injection for
  thermal quench mitigation on {DIII-D},'' {\em Nuclear Fusion}, vol.~56,
  no.~4, p.~046007, 2016.

\bibitem{Shiraki16}
D.~Shiraki, ``Shattered pellet injection as the primary disruption mitigation
  technique for {ITER},'' {\em 26th IAEA fusion energy conference, Kyoto,
  Japan}, 2016.

\bibitem{Shiraki_Commaux16}
D.~Shiraki, N.~Commaux, L.~R. Baylor, N.~W. Eidietis, E.~M. Hollmann, C.~J.
  Lasnier, and R.~A. Moyer, ``Thermal quench mitigation and current quench
  control by injection of mixed species shattered pellets in {DIII-D},'' {\em
  Physics of Plasmas}, vol.~23, no.~6, p.~062516, 2016.

\bibitem{Parks_1978}
P.~B. Parks and R.~J. Turnbull, ``Effect of transonic flow in the ablation
  cloud on the lifetime of a solid hydrogen pellet in a plasma,'' {\em The
  Physics of Fluids}, vol.~21, no.~10, pp.~1735--1741, 1978.

\bibitem{Felber_1979}
F.~Felber, P.~Miller, P.~Parks, R.~Prater, and D.~Vaslow, ``Effects of atomic
  processes on fuel pellet ablation in a thermonuclear plasma,'' {\em Nuclear
  Fusion}, vol.~19, pp.~1061--1072, aug 1979.

\bibitem{Parks_distortion}
P.~B. Parks, ``Magnetic-field distortion near an ablating hydrogen pellet,''
  {\em Nuclear Fusion}, vol.~20, no.~3, p.~311, 1980.

\bibitem{Kuteev_1995}
B.~Kuteev, ``Hydrogen pellet ablation and acceleration by current in high
  temperature plasmas,'' {\em Nuclear Fusion}, vol.~35, no.~4, p.~431, 1995.

\bibitem{Pegourie_HPI2}
F.~Koechl, B.~Pégourié, A.~Matsuyama, H.~Nehme, V.~Waller, D.~Frigione,
  L.~Garzotti, G.~Kamelander, V.~Parail, and J.~E. contributors, ``Modelling of
  pellet particle ablation and deposition: The hydrogen pellet injection code
  hpi2,'' {\em Preprint EFDA–JET–PR(12)57}.

\bibitem{Ishizaki04}
R.~Ishizaki, P.~B. Parks, N.~Nakajima, and M.~Okamoto, ``Two-dimensional
  simulation of pellet ablation with atomic processes,'' {\em Physics of
  Plasmas}, vol.~11, no.~8, pp.~4064--4080, 2004.

\bibitem{Samulyak_2007}
R.~Samulyak, T.~Lu, and P.~Parks, ``A magnetohydrodynamic simulation of pellet
  ablation in the electrostatic approximation,'' {\em Nuclear Fusion}, vol.~47,
  pp.~103--118, Jan 2007.

\bibitem{LuParksSam09}
T.~Lu, P.~Parks, and R.~Samulyak, ``Charging and exb rotation of ablation
  clouds surrounding refueling pellets in hot fusion plasmas,'' {\em Physics of
  Plasmas}, vol.~16, p.~060705, 2009.

\bibitem{Bosviel2020}
N.~Bosviel, P.~Parks, and R.~Samulyak, ``Near-field simulations of pellet
  ablation for disruptions mitigation in tokamaks,'' {\em Physics of Plasmas},
  2020.
\newblock submitted.

\bibitem{Rozhanskij_1994}
V.~Rozhanskij and I.~Veselova, ``Plasma propagation along magnetic field lines
  after pellet injection,'' {\em Nuclear Fusion}, vol.~34, pp.~665--674, may
  1994.

\bibitem{Parks_2000}
P.~B. Parks, W.~D. Sessions, and L.~R. Baylor, ``Radial displacement of pellet
  ablation material in tokamaks due to the grad-b effect,'' {\em Physics of
  Plasmas}, vol.~7, no.~5, pp.~1968--1975, 2000.

\bibitem{Parks_2005}
P.~Parks and L.~Baylor, ``Effect of parallel flows and toroidicity on
  cross-field transport of pellet ablation matter in tokamak plasmas,'' {\em
  Physical Review Letters}, vol.~94, p.~125002, 2005.

\bibitem{Baylor_2000}
L.~Baylor, T.~Jernigan, P.~Gohil, G.~Schmidt, K.~Burrell, S.~Combs, D.~Ernst,
  C.~Greenfield, R.~Groebner, W.~Houlberg, C.~Hsieh, M.~Murakami, P.~Parks,
  M.~Porkolab, W.~Sessions, G.~Staebler, and E.~Synakowski, ``Improved fueling
  and transport barrier formation with pellet injection from different
  locations on diii-d,'' 01 2000.

\bibitem{Lang_1997}
P.~T. Lang, K.~B\"uchl, M.~Kaufmann, R.~S. Lang, V.~Mertens, H.~W. M\"uller,
  and J.~Neuhauser, ``High-efficiency plasma refuelling by pellet injection
  from the magnetic high-field side into asdex upgrade,'' {\em Phys. Rev.
  Lett.}, vol.~79, pp.~1487--1490, Aug 1997.

\bibitem{SamWangChen2018}
R.~{Samulyak}, X.~{Wang}, and H.-C. {Chen}, ``{Lagrangian particle method for
  compressible fluid dynamics},'' {\em Journal of Computational Physics},
  vol.~362, pp.~1--19, June 2018.

\bibitem{Zeldovich}
Y.~Zel'dovich and Y.~Raiser, {\em Physics of shock waves and high temperature
  hydrodynamic phenomena}.
\newblock Dover, 2002.

\bibitem{Parks20}
J.~Zhang and P.~Parks, ``Analytical formula for pellet fuel source density in
  toroidal plasma configuration based on an areal deposition model,'' {\em
  Nuclear Fusion}, apr 2020.

\bibitem{Parks_SCIDAC}
P.~B. Parks, ``On perpendicular conductivity for a partially ionized pellet
  ablation cloud,'' {\em To be submitted}, 2017.

\bibitem{cretin}
H.~A. Scott, ``Cretin—a radiative transfer capability for laboratory
  plasmas,'' {\em Journal of Quantitative Spectroscopy and Radiative Transfer},
  vol.~71, no.~2, pp.~689 -- 701, 2001.
\newblock Radiative Properties of Hot Dense Matter.

\bibitem{BursteddeWilcoxGhattas11}
C.~Burstedde, L.~C. Wilcox, and O.~Ghattas, ``{\texttt{p4est}}: Scalable
  algorithms for parallel adaptive mesh refinement on forests of octrees,''
  {\em SIAM Journal on Scientific Computing}, vol.~33, no.~3, pp.~1103--1133,
  2011.

\bibitem{mu_target}
K.~T.~McDonald et~al., ``The primary target facility for a neutrino factory based on
  muon beams,'' {\em Proc. 2001 Part. Accel. Conf.}, p.~1583, Chicago, IL, June
  2001.

\bibitem{ShihSam19}
W.~Shih, R.~Samulyak, S.~Hsu, S.~Langendorf, K.~Yates, and Y.~C.~F. Thio,
  ``Simulation study of the influence of experimental variations on the
  structure and quality of plasma liners,'' {\em Physics of Plasmas}, vol.~26,
  p.~032704, 2019.

\bibitem{Parks_private_com}
P.~B. Parks, ``The ablation rate of light-element pellets with a kinetic
  treatment for penetration of plasma electrons through the ablation cloud,''
  {\em to be submitted to Physics of Plasmas}, 2020.

\end{thebibliography}

\end{document}